\documentclass[a4paper,twoside,12pt]{report}
\usepackage{amsfonts,a4wide}
\usepackage{amssymb,amsfonts}
\usepackage{amsmath,exscale,amsthm}
\usepackage{latexsym,fancyhdr}
\usepackage{graphicx,psfrag}
\usepackage{cite}
\newcommand{\beq}{\begin{eqnarray}}
\newcommand{\eeq}{\end{eqnarray}}
\newcommand{\be}{\begin{equation}}
\newcommand{\ee}{\ee{equation}}

\newcommand{\R}{\mathbb{R}}
\newcommand{\Ro}{\mathbb{R}^{+}_0}
\newcommand{\f}{\textrm{f}}
\newcommand{\g}{\textrm{g}}
\newcommand{\h}{\textrm{h}}
\newcommand{\ex}{\textrm{e}}

\newcommand{\mjo}{\textrm{J}_0(kR)}
\newcommand{\mjoi}{\textrm{J}_0(k'R)}
\newcommand{\mjI}{\textrm{J}_1(kR)}
\newcommand{\mjIi}{\textrm{J}_1(k'R)}
\newcommand{\mjos}{\textrm{J}_0(kR')}
\newcommand{\mjois}{\textrm{J}_0(k'R')}
\newcommand{\mjIs}{\textrm{J}_1(kR')}
\newcommand{\mjIis}{\textrm{J}_1(k'R')}
\newcommand{\mjII}{\textrm{J}_2(kR)}
\newcommand{\mjIIi}{\textrm{J}_2(k'R)}

\newcommand{\nn}{\nonumber}

\fancypagestyle{plain}{\fancyhf{}}

\newcommand{\clearemptydoublepage}{\newpage{\pagestyle{empty}\cleardoublepage}}

\begin{document}
\bibliographystyle{phaip}
%Titelpage

\begin{titlepage}
\begin{center}

\vspace*{1cm}

{\huge \bf Towards a Gauge Invariant }

\smallskip

{\huge \bf Scattering Theory of}

\smallskip

{\huge \bf Cylindrical Gravitational Waves}

\vspace{3.5cm}{\Large \bf Diplomarbeit}

\vspace{0.5cm} der Philosophisch-naturwissenschaftlichen Fakult\"at

\vspace{0.5cm} der Universit\"at Bern

\vspace{0.5cm} vorgelegt von

\vspace{0.5cm} Andrea Macrina

\vspace{2.5cm} April 2002

\vspace{2.5cm} Leiter der Arbeit:

\vspace{0.5cm} Prof. Petr H\'{a}j\'{\i}\v{c}ek

Institut f\"ur theoretische Physik, Universit\"at Bern

\end{center}
\end{titlepage}
\clearpage

%%% Local Variables: 
%%% mode: latex
%%% TeX-master: "liz"
%%% End: 

\clearemptydoublepage
\setcounter{page}{1}
\pagenumbering{roman}
\tableofcontents
\clearemptydoublepage
%
%Anfang Dokument
\setcounter{page}{1}
\pagenumbering{arabic}
%Introduction

\chapter{Introduction}
One of the most important and unresolved questions of canonical gravity is the
 so called ``problem of time''. In quantum theory time is not regarded as an
 observable in the usual sense, since it is not represented by an
 operator. Time is rather treated as a parameter which, as in classical
 mechanics, describes the evolution of a system. In the time-dependent
 Schr\"odinger equation, for example, time occurs only as parameter and not
 as operator. In the conventional Copenhagen interpretation the notion of a
 measurement made at a \textit{particular} time is a fundamental ingredient.
 An observable is something whose value can be measured at a \textit{fixed} time.\\
\indent In general relativity the role of time is very different. The four
 dimensional spacetime, i.e. a four dimensional manifold with a Lorentzian
 metric and an appropriate topology, can be foliated in many different ways as
 a one-parameter family of spacelike hypersurfaces. Each of these parameters
 can be viewed as a possible definition of time. There are many such
 foliations and there is no way to select a particular one to consider it
 as 'natural'. Hence, in general relativity,
 time is \textit{not a fixed} quantity, as it is in quantum theory. Time is dependent on the chosen foliation of the manifold and is closely linked to a fixed choice of the metric. Such a definition of time does not give any hint how it should be measured, and is so a non-physical quantity.
\\
\indent The gravity group of the institute of theoretical physics deals with
asymptotically flat gravitational models. This kind of systems seems to remove
the ``problem of time'' in the asymptotic, since the parameter of time can
be chosen in a unique manner. The dynamics of the system can be determined, and one
can attempt to study theories like the scattering one in a quantum
theoretical way. The hope is to understand the structure of the theory of such
models in order to be able to state something in the direction of a quantum
gravity.\\
\indent In this diploma thesis such an asymptotically flat model is
 investigated. The model I have chosen in this paper is pure gravity - no
 matter fields are present - with cylindrical symmetry. The
 symmetries of the cylindrical gravitational wave help to solve the Einstein
 equations and thus, to know the metric which describes spacetime of the
 treated toy model. The cylindrically symmetric gravitational wave, also
 called Einstein-Rosen wave, was discovered by A. Einstein and N. Rosen
 \cite{ER37}. To give an idea what gravitational waves are we cite
 \cite{mtw73}: Just as one identifies as water waves small ripples rolling
 across the ocean, so one gives the name gravitational waves to small ripples
 rolling across spacetime.
In the seventies of the last century this field theory was again picked up by
 K. Kucha\v{r} \cite{kuchar71}. The model was solved in the canonical ADM
 formalism and afterwards quantized. Twenty years later, C.G. Torre \cite{torre91} found a complete set of observables of the cylindrically symmetric gravitational wave.\\
\indent This work is mainly based on the contributions mentioned above. In
chapter two the general cylindrical symmetric metric is studied and
mathematical simplifications of it are discussed in detail. Then the
Einstein-Rosen metric is introduced and the corresponding Einstein equations
are derived. In the third chapter, we determine the solution of the
Einstein equations and check the boundary conditions the solving function is
 supposed to meet. In the following
chapter the ADM formalism is presented, firstly for the general spacetime then
for the cylindrically symmetric one and finally for the Einstein-Rosen wave,
which is the interesting case for our consideration. The fifth chapter is
devoted to the derivation of the dynamical variables by solving the Hamilton
equation. Moreover it is shown how
in Torre's work the set of observables for the cylindrical gravitational waves
is obtained. In the last chapter the condition is investigated, which phase
space functions have to fulfill in order to be declared as observables. Then
the Poisson algebra of the observables is studied and quantum theoretical
conclusions to these quantities are stated. In the appendix a detailed
derivation of the Hankel transformation is noted, as well as the
orthonormality relation of the Bessel functions of the first kind. This
relation plays an important role in the mathematical part of this diploma
thesis. Lengthy computations and verifications have also been
displaced into the appendix in order to preserve the overlook while reading.\\ \newline
\textbf{Conventions and notations}:\\
Spacetime has the signature $S=2$, so the signs of the diagonal elements of
 the Lorentzian metric are (-,+,+,+). We use natural units that means
 $\hbar=c=16\pi G=1$. In order to distinguish clearly the
coordinate $R$ from the Ricci scalar $\textrm{R}$, we write throughout the
coordinate in the italic style and the curvature scalar in the Roman
style. The Greek indices run over 0,1,2,3 while
 the Latin indices only over 1,2,3, corresponding to the three space
 coordinates. Derivatives in general are denoted as
follows: $\frac{\partial}{\partial T}\textrm{f}(T,R)\equiv
\textrm{f},^{\phantom{}}_{\,T},\;\frac{\partial^2}{\partial T^2}\text{f}(T,R)\equiv
\textrm{f},^{\phantom{}}_{\,TT}$. In particular the derivative with respect to
 the time $t$ is denoted with a dot, the one with respect to the radius $r$
 with a prime. All these conventions hold unless otherwise specified.

%%% Local Variables: 
%%% mode: latex
%%% TeX-master: "liz"
%%% End: 

\clearemptydoublepage

% 1. Kapitel: Loesung der Einstein-Rosen-Welle, quadrueinsteinrosen.tex
%
\chapter{Structure of the Cylindrical Spacetime}

\section{Line Element with Cylindrical Symmetry}
A method to quantize a gravitational system is to foliate the
four dimensional spacetime in a (3+1)-dimensional general relativity and
subsequently to reduce it to an \textit{equivalent} (1+1)-dimensional model,
which is coupled to a massless scalar field. An example therefore is the
cylindrically symmetric Einstein-Rosen wave. In this section we are going to study the line
element for the mentioned kind of gravitational waves.\\
\indent We start with considering the line element, which defines
cylindrical spacetime, in its most general form. Afterwards we will deduce
from the general form, the line element in the Einstein-Rosen coordinates
through conformal deformations and coordinate transformations. First of all,
we take for convenience the cylindrical coordinate system to describe points in
spacetime, namely the set of variables $(t,r,\varphi,z)$ with $t\in(-\infty,\infty),\; r\in[0,\infty),\;
\varphi\in[0,2\pi)$ and $z\in(-\infty,\infty)$. The most general line
element with cylindrical symmetry reads:
\begin{equation}
 \label{genline}
ds^2=g_{\alpha\beta}\,x^{\alpha}\,x^{\beta}=g_{00}\,dt^2+2\,g_{01}\,dt\,dr+g_{11}\,dr^2+g_{22}\,d\varphi^2+g_{33}\,dz^2,
\end{equation}
with obviously 
\begin{equation}
  \label{cylcoord}
x^0=t,\quad x^1=r,\quad x^2=\varphi,\quad x^3=z.
\end{equation}
To get a cylindrically symmetric line element the coefficients of the metric
$g_{\alpha\beta}$ has to depend only on the coordinates $t$ and $r$. As we are
going to see, only in that case the line element is invariant with respect to
changes of the coordinates $\varphi$ and $z$.
So the metric is a function of the two coordinates $t$ and $r$, and of the form 
\begin{equation}
\label{allgmetric}
g_{\alpha\beta}(t,r)=
\left(\begin{array}{cccc}
g_{00}(t,r) & g_{01}(t,r) & 0 & 0 \\
g_{10}(t,r) & g_{11}(t,r) & 0 & 0 \\
0 & 0 & g_{22}(t,r) & 0 \\
0 & 0 & 0 & g_{33}(t,r)
\end{array}\right),
\end{equation}
where, due to the symmetry property of both indices of the metric tensor $g_{\alpha\beta}$, $g_{01}(t,r)=g_{10}(t,r)$.
After these settings, we now derive the symmetries of spacetime by
analyzing the line element (\ref{genline}). For this inspection we define the $z$-axis as the symmetry axis for
the following four symmetries:
\begin{enumerate}
\item Rotations around the symmetry axis $z$: $\varphi\rightarrow\varphi'=\varphi+\Delta\varphi$.
\item Translations in direction of the $z$-axis: $z\rightarrow z'=z+\Delta z$.
\item Reflections in all surfaces containing the symmetry axis:
  $\varphi\rightarrow\varphi'=-\varphi$.
\item Reflections in all surfaces perpendicular to the z-axis: $z\rightarrow z'=-z$.
\end{enumerate}
If we substitute in the line element $z$ with $z'$ and $\varphi$ with
$\varphi'$ the line element will not change. This is the reason why we have
pretended $g_{\alpha\beta}$ to depend only on $t$ and $r$, otherwise the line
element would change because of the substitutions. Seeing that, $ds^2$ is
invariant under the four transformations noted above, just the four
symmetries. The line element is
symmetric to the following two dimensional transformation group:
\begin{align}  
\label{group}
& \varphi\rightarrow \varphi'=\varphi+\Delta\varphi &\text{(rotation around $z$-axis)}\\
& z\rightarrow z'=z+\Delta z &\text{(translation in direction of the $z$-axis)}
\end{align}
All transformations of such a group are called isometries. Each of these
isometries is generated by the related Killing vector field, for the
cylindrical coordinates $(t,r,\varphi,z)$ the two orthogonal vectors
\begin{eqnarray}
\label{killvec}
\nn \xi ^\alpha_{(\varphi)}&=&(0,0,1,0)\\
    \xi ^\beta_{(z)}&=&(0,0,0,1), 
\end{eqnarray} 
which are determined by the Killing equation
\begin{equation}
\label{killequ}
g_{\alpha\beta,\nu}\,\xi^{\nu}+g_{\nu\beta\,}\xi^{\nu}_{{},\alpha}+g_{\alpha\nu}\,\xi^{\nu}_{{},\beta}=0
\end{equation}
or equivalently
\begin{equation}
\label{orkilling}
\partial_{\kappa}g_{\rho\sigma}=0.
\end{equation}
The two isometries and consequently both Killing fields implicate conservation
laws of the dynamic. Indeed  the cylindrical gravitational fields are characterized by the
existence of a two-parameter Abelian group of motions $^4G_2$ with both, to each other orthogonal, hyper-surface-orthogonal spacelike Killing vectors $\xi ^\nu
_{(\varphi)}$, $\xi ^\nu _{(z)}$.
\section{Simplification of the Cylindrically Symmetric \\ Metric}
In order to obtain solutions of the Einstein equations written for the
general cylindrically symmetric metric (\ref{allgmetric}), we have to simplify
it without loosing physically relevant properties. The method, which in our
case is going to be applied, consists in deforming conformally a part of
the four dimensional cylindrical metric (\ref{allgmetric}), and subsequently
to find a coordinate transformation, which leads into an inertial system
IS. Also we will see that the mentioned coordinate transformation brings the
metric, dependent on the new coordinates, into diagonal form.
\subsection{Conformal Deformation}\label{subsecconfdef}
In the beginning, we would like to start with defining the conformal
deformation in space time as follows:
\begin{equation}
\label{deform}
g_{\alpha\beta}=\Omega^{-2}\;g'_{\alpha\beta}.  
\end{equation}
The function $\Omega=\Omega(t,r)$ is a real and positive scalar function. We consider
now a conformal deformation in two dimensions. In this case the metric
$g_{\lambda\kappa}$, with $\lambda,\kappa=0,1$ labeling $(t,r)$, corresponds to a regular $2\times 2$
matrix. From this it follows that 
\begin{equation}
\label{dim}
g_{\lambda\kappa}\;g^{\lambda\kappa}=\delta^{\lambda}_{\kappa}\;\delta^{\kappa}_{\lambda}=\sum^1_0\,\delta^{\lambda}_{\lambda}=2, 
\end{equation}
where in section 2.2 $\lambda$, $\kappa$, $\rho$, $\sigma$, $\mu$ and $\nu$
run over 0 and 1. Then we are going to show that there is always a conform deforming
factor such that the metric $g_{\lambda\kappa}$ can be expressed through a
flat metric $g'_{\lambda\kappa}$.\\
\\
\textbf{Theorem:} Every two dimensional metric $g_{\lambda\kappa}$ is
conformally flat.\\
\\
\textit{Proof:} We begin with affiliating the transformation law between the Ricci
scalar of a conformally deformed metric, $\textrm{R}[g'_{\lambda\kappa}]$, and the curvature scalar of the
original one, $\textrm{R}[g_{\lambda\kappa}]$. From  the definition of the
conformal deformation (\ref{deform}) it follows:
\begin{equation}
\label{confdeform}
g^{'\lambda\kappa}=\Omega^2\,g^{\lambda\kappa}.  
\end{equation}
The Ricci scalar $\textrm{R}$ is defined by
\begin{equation}
\label{curvscalar}
\textrm{R}=g^{\lambda\kappa}\;\textrm{R}_{\lambda\kappa}=g^{\lambda\kappa}\;\textrm{R}^{\rho}_{\lambda\rho\kappa},  
\end{equation}
where $\textrm{R}_{\lambda\kappa}$ is the Ricci tensor and
$\textrm{R}^{\rho}_{\lambda\rho\kappa}$ the contracted Riemann tensor.
The transformed curvature scalar reads:
\begin{equation}
\label{curvscalarprime}
\textrm{R}^{'\phantom{}}=g^{'\lambda\kappa}\;\textrm{R}^{'\phantom{}}_{\lambda\kappa}=g^{'\lambda\kappa}\;\textrm{R}^{'\rho}_{\phantom{}\lambda\rho\kappa}. 
\end{equation}
If we put in (\ref{curvscalarprime}) instead of $g'_{\,\lambda\kappa}$
consequently $\Omega^2 g^{\lambda\kappa}$ and write down the transformed
Ricci scalar $\textrm{R}'$, we get
\begin{equation}
\label{ricciprime}
\textrm{R}^{'\phantom{}}=\Omega^2\,g^{\lambda\kappa}\,\left[\partial_\rho\Gamma^{'\rho}_{\phantom{x}\lambda\kappa}-\partial_\kappa\Gamma^{'\rho}_{\phantom{x}\lambda\rho}+\Gamma^{'\rho}_{\phantom{x}\nu\rho}\Gamma^{'\nu}_{\phantom{x}\lambda\kappa}-\Gamma^{'\rho}_{\phantom{x}\nu\kappa}\Gamma^{'\nu}_{\phantom{x}\lambda\rho}\right].  
\end{equation}
Evidently, the Christoffel symbol 
\begin{equation}
\label{christoffel}
\Gamma^{\rho}_{\phantom{x}\lambda\kappa}=\tfrac{1}{2}\,g^{\rho\sigma}\,\left[g_{\sigma\kappa,\,\lambda}+g_{\sigma\lambda,\,\kappa}-g_{\lambda\kappa,\,\sigma}\right]  
\end{equation}
has to be transformed. Its transformation law is given by utilizing 
(\ref{confdeform}), as done for the curvature scalar (\ref{curvscalar}).
\begin{equation}
\label{christoffelprime}
\Gamma^{'\rho}_{\phantom{x}\lambda\kappa}=\tfrac{1}{2}\,\Omega^2
g^{\rho\sigma}\,\left[\left(\Omega^{-2}\,g_{\sigma\kappa}\right),_\lambda+\left(\Omega^{-2}\,g_{\sigma\lambda}\right),_\kappa-\left(\Omega^{-2}\,g_{\lambda\kappa}\right),_\sigma\right].  
\end{equation}
If one performs the partial derivatives and the relation
\begin{equation}
\label{derivln}
\ln\,\Omega,_\nu\doteq\left(\ln\,\Omega\right),_\nu=\frac{\Omega,_\nu}{\Omega} 
\end{equation}
is employed, then the transformed Christoffel symbol is obtained:
\begin{equation}
\label{chrisprime}
\Gamma^{'\rho}_{\phantom{x}\lambda\kappa}=\Gamma^{\rho}_{\phantom{x}\lambda\kappa}-\ln\,\Omega,_\lambda\,\delta^{\rho}_\kappa-\ln\,\Omega,_\kappa\,\delta^{\rho}_{\lambda}+\ln\,\Omega,_\sigma\,g_{\,\lambda\kappa}\,g^{\rho\sigma}.
\end{equation}
This intermediate result is introduced in (\ref{ricciprime}) and the
transformation law for the Ricci scalar is given:
\begin{equation}
\label{cstrans}
\textrm{R}'[g'_{\lambda\kappa}]=\Omega^2\,\textrm{R}[g_{\lambda\kappa}]+\Omega^2\left(2\,\partial_\kappa\partial^\kappa\,\ln\,\Omega+2\,\Gamma^{\kappa\rho}_{\phantom{x}\kappa}\,\ln\,\Omega,_\rho\right).  
\end{equation}
To this end we have used the fact that the covariant partial derivative of a scalar
field (scalar function) is a covariant vector field.
\begin{equation}
\label{vectorfield}
V_\rho\doteq\Phi,_\rho=\partial_\rho\,\Phi=\ln\Omega,_\rho.  
\end{equation}
From this it follows the repeatedly applied equation
\begin{eqnarray}
\label{derivvec}
\nn g^{\lambda\kappa}\,\partial_\kappa V_\lambda
\nn &=&g^{\lambda\kappa}\,\left(\partial_\kappa V_\lambda\right)\\
\nn &=&g^{\lambda\kappa}\,\partial_\kappa\,\left(V^{\mu}\,g_{\mu\lambda}\right)\\ 
\nn &=&\delta^\kappa_\mu\,\partial_\kappa
       V^{\mu}+V^{\mu}\,g^{\lambda\kappa}\,g_{\mu\lambda},_\kappa\\
&=&\partial_\kappa V^{\kappa}+V^{\mu}\,g^{\lambda\kappa}\,g_{\mu\lambda},_\kappa. 
\end{eqnarray}
Further more the subsequent expression is  relevant also:
\begin{equation}
\label{gammarel}
2\,V^\rho\,\Gamma^{\lambda}_{\phantom{x}\rho\lambda}=g^{\lambda\kappa}\,g_{\lambda\kappa},_\rho\,V^{\rho}.  
\end{equation}
In case that $\Omega$ is a scalar field, the right hand side of the equation
(\ref{cstrans}) can be noted more compactly. So, we apply the
(two dimensional) Laplacian operator $\triangle$ onto a scalar field in the
two dimensional spacetime. 
\begin{eqnarray}
\label{laplace}
\nn \triangle\phi&=&\nabla_{\kappa}\left(g^{\lambda\kappa}\,\nabla_{\lambda}\,\phi\right)\\
\nn &=& \nabla_{\kappa}V^\kappa\\
\nn &=& \partial_{\kappa}V^\kappa+\Gamma^{\kappa}_{\phantom{x}\rho\kappa}V^{\rho}\\
    &=& \partial_{\kappa}\partial^{\kappa}\Phi+\Gamma^{\kappa}_{\phantom{x}\rho\kappa}\,\partial^{\rho}\Phi,  
\end{eqnarray}
whereby we know that the covariant derivative of a scalar field is equal to
the covariant partial derivative of the same. From (\ref{vectorfield}) we see
that $\Phi=\ln\Omega$, and by putting it into (\ref{cstrans}), we achieve that
the transformation law for the Ricci scalar reduces to the following compact 
formula:
\begin{equation}
\label{riccitrans}
\textrm{R}'[g'_{\lambda\kappa}]=\Omega^2\,\textrm{R}[g_{\lambda\kappa}]+2\,\Omega^2\,\triangle\ln\Omega.
\end{equation}
Now we set the two dimensional conformally deformed metric
$g'_{\lambda\kappa}$ to be flat. In this case the transformed Ricci scalar
$\textrm{R}'[g'_{\lambda\kappa}]$ vanishes. From the transformation law
(\ref{riccitrans}) it follows then:
\begin{equation}
\label{diffeq}
\triangle\ln\Omega+\tfrac{1}{2}\,\textrm{R}[g_{\lambda\kappa}]=0  
\end{equation}
This equation is the known generalized inhomogeneous Klein-Gordon
differential equation for the massless scalar field $\Phi=\ln\Omega$. This
kind of differential equation can be worked out by performing a change of
variables to the double null coordinates:
\begin{equation}
  \label{doubcoord}
u\doteq t+r,\qquad v\doteq t-r.   
\end{equation}
In \cite{kuchar71} it is shown that by the substitution to the double null coordinates the differential
equation (\ref{diffeq}) possesses for all regular $\Omega>0$ a solution. So it
exists throughout a scalar function $\Omega$ such that the Klein-Gordon
equation is fulfilled. Thus, there is a conformal deforming scalar function $\Omega$ in such a way that the metric
$g_{\lambda\kappa}$ is expressed by a flat metric $g'_{\lambda\kappa}$. 
\subsection{Coordinate Transformation into the Inertial System}\label{subsecIS}
In the previous paragraph it has been showed that the $(t,r)$-part of the
cylindrically symmetric line element is conformally flat. By virtue of this
property it can be expressed by a two dimensional flat metric.
\begin{equation}
\label{twodimline}
g_{\lambda\kappa}\,dx^{\lambda}\,dx^{\kappa}=\Omega^{-2}\,g'_{\lambda\kappa}\,dx^{\lambda}\,dx^{\kappa},  
\end{equation}
\begin{equation}
\label{twodimcomps}
g'_{00}=\Omega^{-2}\,g_{00},\qquad g'_{01}=\Omega^{-2}\,g_{01},\qquad g'_{11}=\Omega^{-2}\,g_{11}. 
\end{equation}
The scalar function is chosen such that $R'[g'_{\lambda\kappa}]=0$ holds, in
order the metric $g'_{\lambda\kappa}$ to be flat.
But by then, the four dimensional metric (\ref{allgmetric}) has not been
simplified really, since the number of its non-vanishing components did not
decrease. Such a simplification would be an advantage to find solutions to
the vacuum Einstein equations, as the number of vanishing components would
rise. At this place it is again important to remind that the physical model
has not to be trivialized by the possible elimination of some components of
the Einstein equations. The issue is only to obviate  the mathematical
impediments by a suitable choice of coordinates.\\
\indent The theorem about the conformal flatness of two dimensional metrics
just allow us to find a global coordinate transformation to an
inertial system, as the conformally flat metric can be described by a flat
metric $g'_{\lambda\kappa}$. Now the importance of the proved theorem appears
evidently on the score of the suited simplification of the Einstein equations
by reducing the number of non-vanishing components of the four dimensional
metric (\ref{allgmetric}). In the coordinates $\bar{t}$ and $\bar{r}$ of the inertial system the
conformally deformed two dimensional $(t,r)$-part of (\ref{allgmetric}),
namely $g'_{\lambda\kappa}$, just transmute to the Minkowskian standard metric:
\begin{align}
\label{deformminkmetric}
&g'_{\lambda\kappa}(t,r)= 
\left(\begin{array}{cc}
\frac{g_{00}}{\Omega^{2}} & \frac{g_{01}}{\Omega^{2}}\\
\frac{g_{01}}{\Omega^{2}} & \frac{g_{01}}{\Omega^{2}}
\end{array}\right) &\xrightarrow{\text{coordinate transformation}\ \varphi}&
&\eta_{\lambda\kappa}(\bar{t},\bar{r})= 
\left(\begin{array}{cc}
-1 & 0\\
0 & 1
\end{array}\right)
\end{align}
Through the conformal deformation $\Omega$ and the coordinate transformation $\varphi$ we get from the conformally flat metric
$g_{\lambda\kappa}(t,r)$ a diagonal matrix, namely
\begin{align}
\label{diag}
&g_{\lambda\kappa}(t,r) &\xrightarrow{\hspace{1cm}\Omega,\ \varphi\hspace{1cm}}& 
&\bar{g}_{\lambda\kappa}(\bar{t},\bar{r})=\bar{\Omega}^2(\bar{t},\bar{r})\,\eta_{\lambda\kappa}(\bar{t},\bar{r})=
\left(\begin{array}{cc}
-\bar{\Omega}^2 & 0\\
0 & \bar{\Omega}^2
\end{array}\right).  
\end{align}
The conformal deformation and the coordinate transformation simplify the
exclusively on $t$ and $r$ depending four dimensional metric
(\ref{allgmetric}) significantly, as it is brought into diagonal form without
changing anything on the physical aspect. The four dimensional metric reads
newly:
\begin{equation}
\label{barmetric}
\bar{g}_{\alpha\beta}(\bar{t},\bar{r})=
\left(\begin{array}{cccc}
-\bar{\Omega}^2 & 0 & 0 & 0 \\
0 & \bar{\Omega}^2 & 0 & 0 \\
0 & 0 & g_{22}(\bar{t},\bar{r}) & 0 \\
0 & 0 & 0 & g_{33}(\bar{t},\bar{r})
\end{array}\right),
\end{equation}
with the knowledge that $g_{22}(\bar{t},\bar{r})$ and
$g_{33}(\bar{t},\bar{r})$ are invariant with respect to the coordinate
transformation. This fact is shown to hold in equation (\ref{fcts}) if we are
allowed to anticipate $g_{22}(\bar{t},\bar{r})=R^2\,\ex^{-\psi}$ and
$g_{33}(\bar{t},\bar{r})=\ex^{\psi}$ in case of the Einstein-Rosen spacetime.\\
\indent At last we would like to add an illustration for a better intuitively
understanding of the coordinate transformation into the inertial system. The
starting point is a flat geometry, which is represented by the so-called flat
metric $g'_{\lambda\kappa}$ in (\ref{deformminkmetric}). However we probably would not assert on the basis of the
treated metric that a two dimensional flat geometry is described, since for
this case we rather would expect the Minkowskian standard
metric. Nevertheless, we get zero by calculating the Ricci scalar of
$g'_{\lambda\kappa}$; then we say the metric is flat. How should we understand
intuitively this apparent paradox? To this end, we consider a flat disk at rest
in an inertial system. With respect to the coordinates $(\bar{t},\bar{r})$ of
an inertial system, an observer A measures the length $\textrm{C}=2\pi$ for the
circumference of the disk with the unit radius. Now we make the assumption
that an other observer B sizes the disk with respect to a coordinate system
rotating around the symmetry axis, which intersects the middle point of the
disk. We emphasize that the rotating coordinate system is accelerated. The
observer B also measures the radius to be unit, but a circumference $\textrm{C}'$
smaller than $2\pi$. This is an effect due to the Lorentz contraction. So the
observer B, confident in his ratios, would not assert the measured surface to
be a disk, but rather a paraboloid. Thus, he would confirm a curved geometry
and consequently a two dimensional metric, which is different from the Minkowskian standard
metric, presumably with four non-vanishing components. However the Ricci
scalar of his metric is zero implicating a flat geometry.\\
In a certain sense, he has applied for his gaging the ``wrong'' coordinate
system, namely an accelerated one, which intuitively speaking, distorts the
geometry. So it is recommendable to undertake a coordinate transformation to
the inertial system to see the geometry the ``right'' way. That the Ricci scalar of a flat
metric vanishes with respect to all coordinate systems is not surprisingly,
since a scalar is independent on coordinates. That is the reason why also
observer B has to get a vanishing Ricci scalar for his metric.\\
\indent By means of this Gedankenexperiment, we see that the physical
model is not influenced at all by a coordinate transformation. In fact the
properties of the two dimensional geometry are not changed, it remains flat
with respect to all coordinate systems.       
\section{Einstein-Rosen Metric}\label{secERM}
After the introductory sections, we shall consider more concretely the real subject of our
study, namely the Einstein-Rosen gravitational waves. The utilized metric for
this model is the Einstein-Rosen metric, namely
\begin
{equation}
\label{ermetric}
g^{ER}_{\alpha\beta}(T,R)=
\left(\begin{array}{cccc}
-\ex^{\Gamma-\psi} & 0 & 0 & 0 \\
0 & \ex^{\Gamma-\psi} & 0 & 0 \\
0 & 0 & R^2\;\ex^{-\psi} & 0 \\
0 & 0 & 0 & \ex^{\psi}
\end{array}\right).
\end{equation}
The functions $\Gamma$ and $\psi$ are dependent on both Einstein-Rosen
coordinates $T$ and $R$ exclusively. The complete coordinate system of course
is eked by the other two spatial coordinates $\varphi$ and $z$. As we notice,
the Einstein-Rosen metric is cylindrically symmetric, since it is a special
case of the general cylindrically symmetric metric (\ref{allgmetric}). The
metric (\ref{ermetric}) depends on the time coordinate
$T\in[0,\infty)$ and on the spatial coordinate $R\in[0,\infty)$ only. Just
this requirement was the second requisite in section 2.1 a metric is
supposed to possess in order to be cylindrically symmetric. Thus, the
Einstein-Rosen metric inherits all properties of cylindrical metrics and which
are listed in section 1.1. \\
\indent In the following it is showed, how one gets from the general
cylindrically symmetric metric (\ref{allgmetric}) the Einstein-Rosen metric,
which is the important special case for our purpose. If we set in the general
metric (\ref{allgmetric})
\begin{align}
\label{gab}
&g_{22}(t,r)=R^2\,\ex^{-\psi}, &g_{33}(t,r)=\ex^{\psi},   
\end{align}
\footnote{Compare with equation (2) in section II of \cite{kuchar71}}we remark that only the two dimensional $(t,r)$-part
\begin{equation}
\label{twometric}
g_{\lambda\kappa}(t,r)=
\left(\begin{array}{cc}
g_{00}(t,r) & g_{01}(t,r)\\
g_{01}(t,r) & g_{11}(t,r)
\end{array}\right) 
\end{equation}
has to be brought into the diagonal form
\begin{equation}
\label{erdiag}
g^{ER}_{\lambda\kappa}(T,R)=
\left(\begin{array}{cc}
-\ex^{\Gamma-\psi} & 0\\
0 & \ex^{\Gamma-\psi}
\end{array}\right)=
\ex^{\Gamma-\psi}\,
\left(\begin{array}{cc}
-1 & 0\\
0 & 1
\end{array}\right),  
\end{equation}
where $\lambda$ and $\kappa$ run again over 0 and 1.
In the original coordinates $(t,r,\varphi,z)$ the functions $R$ and $\psi$ are
dependent on $t$ and $r$. In Kucha\v{r}'s paper \cite{kuchar71} both functions are
defined by the Killing vectors (\ref{killvec}):
\begin{align}
\label{scalarR}
&R(t,r)\doteq\sqrt{\xi^\alpha_{(\varphi)}\
  \xi^{\phantom{\alpha}}_{{\alpha(\varphi)}}\ \xi^\beta_{(z)}\
  \xi^{\phantom{\beta}}_{{\beta(z)}}}\\
\label{scalarpsi}
&\psi(t,r)\doteq\ln(\xi^\alpha_{(z)}\ \xi^{\phantom{\alpha}}_{{\alpha(z)}}).  
\end{align}
If the right hand side of both equations (\ref{scalarR}) and (\ref{scalarpsi})
are calculated by means of both Killing vectors and the metric 
\begin{equation}
\label{qbarmetric}
g_{\alpha\beta}(t,r)=
\left(\begin{array}{cccc}
g_{00} & g_{01} & 0 & 0 \\
g_{01} & g_{11} & 0 & 0 \\
0 & 0 & R^2\;\ex^{-\psi} & 0 \\
0 & 0 & 0 & \ex^{\psi}
\end{array}\right),  
\end{equation}
then the functions are got. Now we bring $g_{\lambda\kappa}(t,r)$ (\ref{twometric}) into
diagonal form. The metric is two dimensional and dependent only on $t$ and
$r$. After we have learned in section 2.2 how this kind of metric can be
simplified, the following plan appears to be suited:
\begin{enumerate}
\item We find a scalar function $\Omega$ such that the metric
  $g_{\lambda\kappa}$ can be expressed by a flat metric
  $g'_{\lambda\kappa}$. To this end the Klein-Gordon differential equation for
  the massless scalar field $\Omega$ (\ref{diffeq}) must be solved.
\item By a general coordinate transformation $\varphi$, a coordinate system
  $(\bar{t},\bar{r})$ of an inertial system IS is found, wherein the flat
  metric $g'_{\lambda\kappa}(\bar{t},\bar{r})$ assumes the Minkowskian
  standard form.
\end{enumerate}
The result of this procedure is the metric which is conformally flat in the
$(t,r)$-part. This metric is noted in the second section of Kucha\v{r}'s paper
\cite{kuchar71},   
\begin{equation}
\label{barmetrik}
\bar{g}_{\alpha\beta}(\bar{t},\bar{r})=
\left(\begin{array}{cccc}
-\ex^{\bar{\gamma}-\psi} & 0 & 0 & 0 \\
0 & \ex^{\bar{\gamma}-\psi} & 0 & 0 \\
0 & 0 & R^2\;\ex^{-\psi} & 0 \\
0 & 0 & 0 & \ex^{\psi}
\end{array}\right),
\end{equation}
for the case that the solution of the Klein-Gordon differential equation (\ref{diffeq}) is
the subsequently conform deforming factor:
\begin{eqnarray}
\label{conffactor}
&\Omega^2=\ex^{\bar{\gamma}-\psi},&\\
\nn\\
\label{fcts}
&\bar{\gamma}=\bar{\gamma}(\bar{t},\bar{r}),\quad\psi=\bar{\psi}(\bar{t},\bar{r})=\psi(t,r).&  
\end{eqnarray}
The functions $R$ and $\psi$ do not change because of the coordinate
transformation $\varphi$, since they behave like scalars\footnote{Now the
  provision we made for the metric (\ref{barmetric}) is justifiable
  in virtue of the behavior of $R$ and $\psi$ with respect to coordinate transformations.}, as one can take from
the equations (\ref{scalarR}) and (\ref{scalarpsi}).\\
\indent We shortly focus on the Einstein-Rosen coordinate $R$. To develop a more concrete vision, as also to convince us that the
Einstein-Rosen coordinate is a kind of radius in the cylindrical space time,
we add a short illustration. We concentrate on a two-dimensional cylindrical
surface $t$=const., $r$=const. around the axis of symmetry. Let be the height
of the cylindrical surface $\Delta z=1$. Thus the area of the described surface is
$2\pi R$, the same as the area for the surface of a cylinder with height one and
radius $R$ in Euclidean space. So we can identify $R$ in (\ref{barmetric}) with a kind of radius 
in the cylindrical spacetime. To make sure that this quantity is spacelike, a
special regard should be given to the vector $R,_{\nu}(t,r)$, which
can be spacelike ($R,_{\nu}R,^{\nu}>0$), timelike ($R,_{\nu}R,^{\nu}<0$) or
lightlike ($R,_{\nu}R,^{\nu}=0$). In order to deal with the model of
cylindrically symmetric gravitational waves, we are forced to choose the
vector $R,_{\nu}(t,r)$ spacelike everywhere in spacetime. Of course, it
would be quite senseless to speak of a timelike or a lightlike radius vector,
at least if the coordinate $R$ is supposed to form with the timelike
Einstein-Rosen coordinate $T$ an orthogonal coordinate system.
\subsection{Conformal Coordinate Transformation}\label{subsecCCT}
Comparing the metric $\bar{g}_{\alpha\beta}(\bar{t},\bar{r})$ with the
Einstein-Rosen metric $g^{ER}_{\alpha\beta}(T,R)$, we notice that a further
simplification of the first metric is possible. By a coordinate transformation
$\vartheta$ into the coordinate system $(T,R,\varphi,z)$, the function
$R(\bar{t},\bar{r})$ can be converted into a coordinate, namely one of the
Einstein-Rosen coordinates. Looking at both metrics
above, we remark that besides a further conformal deformation in the $(t,r)$-part
\begin{equation}
\label{barconfdeform}
\bar{g}_{\lambda\kappa}=\bar{\Omega}^{-2}\,g^{ER}_{\lambda\kappa},  
\end{equation}
just a change of coordinates transports the one metric in the other (please
notice $\lambda,\kappa=0,1$). Then the
functions $T(\bar{t},\bar{r})$ and $R(\bar{t},\bar{r})$ adopt the role of coordinates $T$ and $R$ in the orthogonal Einstein-Rosen
coordinate system $(T,R,\varphi,z)$. While in these coordinates the function
$\psi(t,r)$ will not change as it is a scalar, the function $\gamma(\bar{t},\bar{r})$, on the
contrary, will change due to the coordinate transformation, according to the
relation,
\begin{equation}
  \label{gG}
\gamma(\bar{t},\bar{r})\rightarrow\Gamma(T,R)=\gamma-\ln\left(R'{^2}-T'{^2}\right),  
\end{equation}
where the primes are in this case derivatives with respect to $\bar{r}$.
Since the change of coordinates influences the time and the
radius, we require the following connection between the $(t,r)$-parts of the
considered metrics:
\begin{equation}
\label{conftrans}
-d\bar{t}^2+d\bar{r}^2=\bar{\Omega}^{-2}\,(-dT^2+dR^2).
\end{equation}
Thereby it is not about a general coordinate transformation, but about a so-called
\textit{conformal} coordinate transformation. From this requirement we are going to get
an additional condition on the coordinates $T$ and $R$. We compute the total
differential of the functions $T(\bar{t},\bar{r})$ and $R(\bar{t},\bar{r})$
and set them in the right hand side of equation (\ref{conftrans}) after
multiplying with $\bar{\Omega}^2$. So we get:
\begin{eqnarray}
\label{cteq}
\nn\bar{\Omega}^{2}\,(-d\bar{t}^2+d\bar{r}^2)=&-&(T,_{\bar{t}}^2-R,_{\bar{t}}^2)d\bar{t}+(R,_{\bar{r}}^2-T,_{\bar{r}}^2)d\bar{r}^2\\
&-&T,_{\bar{t}}\,T,_{\bar{r}}\,d\bar{t}d\bar{r}+R,_{\bar{t}}\,R,_{\bar{r}}\,d\bar{t}d\bar{r}.
\end{eqnarray}
The equation (\ref{cteq}) holds only if one of the following two settings is applied:
\begin{align}
\label{trconnex1}
T,_{\bar{t}}& =R,_{\bar{r}},& T,_{\bar{r}}& =R,_{\bar{t}}\\
\label{trconnex2}
T,_{\bar{t}}& =-R,_{\bar{r}},& T,_{\bar{r}}& =-R,_{\bar{t}}. 
\end{align}
We have assumed that the order of the partial derivatives is commutable, what
for coordinate transformations surely apply.
Derivating partially the first equation in (\ref{trconnex1}) by $\bar{t}$ and
the second in (\ref{trconnex1}) by $\bar{r}$, and then subtracting the second
result from the first, we get
\begin{equation}
\label{boxT}
\Box\,T(\bar{t},\bar{r})\doteq T,_{\bar{t}\bar{t}}-T,_{\bar{r}\bar{r}}=0.  
\end{equation}
Derivating partially the first equation in (\ref{trconnex1}) by $\bar{r}$ and
the second in (\ref{trconnex1}) by $\bar{t}$, and then subtracting the first
result from the second, we get 
\begin{equation}
\label{boxR}
\Box\,R(\bar{t},\bar{r})\doteq R,_{\bar{t}\bar{t}}-R,_{\bar{r}\bar{r}}=0.  
\end{equation}
So, the Einstein-Rosen coordinates $T$ and $R$ not only build by construction an orthogonal
system and are timelike respectively spacelike, but also they have to be
harmonically with respect to the coordinates $\bar{t}$ and $\bar{r}$. Due to
the invariance of $R$ regarding coordinate transformations, they are
harmonically with respect to the original coordinates $t$ and $r$, too.
Of course the coordinate $T$ is also a scalar and therefore invariant to an
arbitrary change of coordinates, as the quantity follows by
integrating the equations (\ref{trconnex1}) or (\ref{trconnex2}).\\
\indent In order to verify whether the function $R(\bar{t},\bar{r})$ is
harmonic, we write the Einstein equations for the metric (\ref{barmetrik}):
\begin{equation}
\label{einsteq}
\bar{\textrm{G}}_{\alpha\beta}=\bar{\textrm{R}}_{\alpha\beta}-\tfrac{1}{2}\,\bar{\textrm{R}}\,\bar{g}_{\alpha\beta}=0  
\end{equation}
By performing the Ricci tensor $\bar{\textrm{R}}_{\alpha\beta}$ and the
Ricci scalar\footnote{In the subsequent section the way how
the tensor and the scalar are performed is explained in a more detailed way.} $\bar{\textrm{R}}$ for the
metric $\bar{g}_{\alpha\beta}$, we obtain the following, for our consideration relevant, components:
\begin{align}
\label{g22}
\bar{\textrm{G}}_{22}&
=R(\bar{\gamma},_{\bar{t}\bar{t}}-\bar{\gamma},_{\bar{r}\bar{r}})+\tfrac{1}{2}\,R(\psi,_{\bar{t}}^2-\psi,_{\bar{r}}^2)=0\\
\label{g33}
\bar{\textrm{G}}_{33}&
=R(\bar{\gamma},_{\bar{t}\bar{t}}-\bar{\gamma},_{\bar{r}\bar{r}})+\tfrac{1}{2}\,R(\psi,_{\bar{t}}^2-\psi,_{\bar{r}}^2)+2(R,_{\bar{t}\bar{t}}-R,_{\bar{r}\bar{r}})=0 
\end{align}
Equation (\ref{boxR}) just appears by subtracting $\bar{\textrm{G}}_{33}$ from
$\bar{\textrm{G}}_{22}$,
\begin{equation}
\label{harmR}
R,_{\bar{t}\bar{t}}-R,_{\bar{r}\bar{r}}=0.  
\end{equation}
We assume the derivative of the Einstein-Rosen coordinate $R$ to be by
construction a spacelike vector $(R,_\nu R,^{\nu}>0)$ and the coordinate itself
to be perpendicularly on the time coordinate $T$. Then $R$ is fit to be
Einstein coordinate, since it is showed it to be harmonic with respect as well to the
variables $(\bar{t},\bar{r})$ as to $(t,r)$.
At last we would like to note both Einstein-Rosen coordinates $T$ and $R$. To
this end we solve both D'Alembert equations (\ref{boxT}) and (\ref{boxR}) for
one time dimension and one space dimension. This sort of differential equation
can be worked out by writing them in double null coordinates, i.e. the
following substitution of variables is suggested:
\begin{align}
\label{uv}
&\bar{u}=\bar{t}+\bar{r}\\
&\bar{v}=\bar{t}-\bar{r}. 
\end{align}
If we note the differential operators in terms of the new coordinates, then we
get for the time and radius derivatives
\begin{align}
\label{uvdiff}
&\frac{\partial}{\partial\bar{t}}=\frac{\partial\bar{u}}{\partial\bar{t}}\frac{\partial}{\partial\bar{u}}+\frac{\partial\bar{v}}{\partial\bar{t}}\frac{\partial}{\partial{\bar{v}}}=\frac{\partial}{\partial\bar{u}}+\frac{\partial}{\partial\bar{v}},\\
\nn\\
&\frac{\partial}{\partial\bar{r}}=\frac{\partial\bar{u}}{\partial\bar{r}}\frac{\partial}{\partial\bar{u}}+\frac{\partial\bar{v}}{\partial\bar{r}}\frac{\partial}{\partial{\bar{v}}}=\frac{\partial}{\partial\bar{u}}-\frac{\partial}{\partial\bar{v}}, 
\end{align}
and putting them in (\ref{boxT}) and (\ref{boxR}), they yield the D'Alembert
differential equations in the double null coordinates:
\begin{align}
\label{uvT}
&\frac{\partial^2}{\partial\bar{u}\partial\bar{v}}\,T(\bar{u},\bar{v})=0,\\
\label{uvR}
&\frac{\partial^2}{\partial\bar{u}\partial\bar{v}}\,R(\bar{u},\bar{v})=0.
\end{align}
So, the general solution of the differential equation (\ref{uvR}) can be
found. For the derivative with respect to $\bar{v}$ the following is compelled
to be set:
\begin{equation}
\label{vR}
\frac{\partial}{\partial\bar{v}}\,R(\bar{u},\bar{v})=\textrm{f}(\bar{v}).  
\end{equation}
Only in case that the derivative with respect to $\bar{v}$ yields a function
depending on $\bar{v}$ exclusively, the following derivative with respect to
$\bar{u}$ vanishes. In that case the D'Alembert equation (\ref{uvR}) is
fulfilled. The indefinite integration over $\bar{v}$ of equation (\ref{vR})
yields the general form of the solution for the $R$-coordinate:
\begin{align}
\label{intvR}
\nn&\int d\bar{v}\frac{\partial}{\partial\bar{v}}\,R(\bar{u},\bar{v})=\int
d\bar{v}\,\textrm{f}(\bar{v})\\
\nn\\
\rightarrow\quad&R(\bar{u},\bar{v})=\textrm{F}(\bar{v})+\textrm{G}(\bar{u}). 
\end{align}
Analogously the general solution for the time coordinate $T$ reads:
\begin{align}
\label{intvT}
\nn&\int d\bar{v}\frac{\partial}{\partial\bar{v}}\,R(\bar{u},\bar{v})=\int
d\bar{v}\,\textrm{h}(\bar{v})\\
\nn\\
\rightarrow\quad&T(\bar{u},\bar{v})=\textrm{H}(\bar{v})+\textrm{K}(\bar{u}). 
\end{align}
As we know, both Einstein-Rosen coordinates are linked to each other by the
relations (\ref{trconnex1}) or (\ref{trconnex2}). From the first identity in
(\ref{trconnex1}) the connection in double null variables follows:
\begin{equation}
\label{truv1}
T,_{\bar{t}}=R,_{\bar{r}}\ \xrightarrow{\quad}\ \frac{\partial\textrm{G}}{\partial\bar{u}}+\frac{\partial\textrm{F}}{\partial\bar{v}}=\frac{\partial\textrm{K}}{\partial\bar{u}}+\frac{\partial\textrm{H}}{\partial\bar{v}}.  
\end{equation}
From the second identity in (\ref{trconnex1}) we get a similar expression performing in the same way
\begin{equation}
\label{truv2}
T,_{\bar{r}}=R,_{\bar{t}}\ \xrightarrow{\quad}\ \frac{\partial\textrm{G}}{\partial\bar{u}}-\frac{\partial\textrm{F}}{\partial\bar{v}}=\frac{\partial\textrm{K}}{\partial\bar{u}}+\frac{\partial\textrm{H}}{\partial\bar{v}}.  
\end{equation}
Adding both identities written in the double null coordinates we obtain 
\begin{eqnarray}
\label{GK}
\nn\frac{\partial\textrm{G}}{\partial\bar{u}}&=&\frac{\partial\textrm{K}}{\partial\bar{u}},\\  
\nn\\
\rightarrow\ \textrm{G}(\bar{u})&=&\textrm{K}(\bar{u}),
\end{eqnarray}
and by putting this result in (\ref{truv1}) or (\ref{truv2}) trivially
\begin{eqnarray}
\label{FH}
\nn\frac{\partial\textrm{F}}{\partial\bar{v}}&=&\frac{\partial\textrm{H}}{\partial\bar{v}}\\
\nn\\
\rightarrow\ \textrm{F}(\bar{v})&=&-\textrm{H}(\bar{v}).  
\end{eqnarray}
We remind on the fact that in (\ref{GK}) and (\ref{FH}) functions depending on
$\bar{u}$ respectively $\bar{v}$ are not allowed to occur due to the indefinite
integrations. Otherwise the condition 
\begin{align}
\label{cond}
&\frac{\partial}{\partial\bar{u}}\frac{\partial}{\partial\bar{v}}\,R(\bar{u},\bar{v})=\frac{\partial}{\partial\bar{u}}\,\textrm{f}(\bar{v})=0,\\
&\textrm{f}(\bar{v})=\frac{\partial}{\partial\bar{v}}\,\textrm{F}(\bar{v}), 
\end{align}
would be violated. Replacing the double null variables by the original
coordinates $\bar{t}$ and $\bar{r}$ the Einstein-Rosen coordinates read:
\begin{align}
\label{TR}
&T(\bar{t},\bar{r})=\textrm{G}(\bar{t}+\bar{r})-\textrm{F}(\bar{t}-\bar{r}),\\
&R(\bar{t},\bar{r})=\textrm{G}(\bar{t}+\bar{r})+\textrm{F}(\bar{t}-\bar{r}). 
\end{align}
As we have seen, the conditions (\ref{trconnex1}) and (\ref{trconnex2}), which
arise from the crucial requirement the transformation between
$(\bar{t},\bar{r})$ and $(T,R)$ to be conformal, is authoritative. If we
construct the coordinate $R(\bar{t},\bar{r})$ then the coordinate
$T(\bar{t},\bar{r})$ emerge directly from the relations (\ref{trconnex1}) and
(\ref{trconnex2}). It has been showed that the function $R(\bar{t},\bar{r})$
is actually harmonic by solving the Einstein equations written for the metric
(\ref{barmetric}). Rather due to the relations (\ref{trconnex1}) and
(\ref{trconnex2}) it is also proved that the time coordinate
$T(\bar{t},\bar{r})$, which corresponds to the Einstein-Rosen coordinate
$R(\bar{t},\bar{r})$, is also harmonic. Therewith the differential equation
(\ref{boxT}) is fulfilled, too.
\section{Einstein Equations in the Einstein-Rosen\\Coordinates }
The model of the cylindrically symmetric gravitational waves does not include
the presence of matter fields. Thus, we consider a general relativity without
matter fields, that means a vacuum spacetime. So, we are interested in the
vacuum Einstein equations
\begin{equation}
\label{einsteineqs}
\textrm{G}_{\alpha\beta}=\textrm{R}_{\alpha\beta}-\tfrac{1}{2}\;\textrm{R}\;g_{\alpha\beta}=0.  
\end{equation}
In this case the energy-momentum tensor $\textrm{T}_{\alpha\beta}$ is zero and
the Einstein equations \\
$\textrm{G}_{\alpha\beta}-\kappa\textrm{T}_{\alpha\beta}=0$ reduce to the
vacuum equations (\ref{einsteineqs}).\\ 
\indent We now would like to calculate the vacuum equations
(\ref{einsteineqs}) in the Einstein-Rosen coordinates system, since, as we are
going to see, they assume a very convenient form therein. So, the metric
$g_{\alpha\beta}$ is obviously the Einstein-Rosen metric (\ref{ermetric}). In virtue of
the cylindrical symmetry, only six components of the Einstein tensor
(\ref{einsteineqs}) do not vanish, namely:
\begin{align}
\label{etcomps}
\textrm{G}_{00}& =\textrm{R}_{00}-\tfrac{1}{2}\;\textrm{R}\;g_{00}&
\textrm{G}_{01}& =\textrm{R}_{01}\\
\textrm{G}_{10}& =\textrm{R}_{10}&
\textrm{G}_{11}& =\textrm{R}_{11}-\tfrac{1}{2}\;\textrm{R}\;g_{11}\\
\textrm{G}_{22}& =\textrm{R}_{22}-\tfrac{1}{2}\;\textrm{R}\;g_{22}&  
\textrm{G}_{33}& =\textrm{R}_{33}-\tfrac{1}{2}\;\textrm{R}\;g_{33}.
\end{align}
Because of the symmetry of the Ricci tensor $\textrm{R}_{\alpha\beta}$ in its
indices, the components $\textrm{G}_{01}$ and $\textrm{G}_{10}$ are identical,
and so one of them can be neglected, for instant $\textrm{G}_{10}$. Now only
five components are relevant for the further computation of the Einstein field
equations. In the apposite appendix C the various non vanishing components of
the Ricci tensor 
\begin{equation}
\label{riccitens}
\textrm{R}_{\alpha\beta}=\textrm{R}^\mu_{\alpha\mu\beta}=\partial_\mu\;\Gamma^\mu_{\alpha\beta}-\partial_\beta\;\Gamma^\mu_{\alpha\mu}+\Gamma^\mu_{\rho\mu}\;\Gamma^\rho_{\alpha\beta}-\Gamma^\mu_{\rho\beta}\;\Gamma^\rho_{\alpha\mu}
\end{equation}
already have been performed, as well as the Ricci scalar
\begin{equation}
\label{riccisc}
 \textrm{R}=g^{\alpha\beta}\;\textrm{R}_{\alpha\beta}. 
\end{equation}
Then the five Einstein equations read:
\begin{align}
\label{eseqs}
\nn \textrm{G}_{00}&= \tfrac{1}{2}\left(\psi,^2_T+\psi,^2_R\right)-\frac{1}{R}\Gamma,_R=0\\
\nn \textrm{G}_{01}&= \psi,_T\,\psi,_R-\frac{1}{R}\Gamma,_T=0\\
\nn \textrm{G}_{11}&= \tfrac{1}{2}\left(\psi,^2_T+\psi,^2_R\right)-\frac{1}{R}\Gamma,_R=0\\
\nn \textrm{G}_{22}&=
\nn -\Gamma,_{RR}-\Gamma,_{TT}-\tfrac{1}{2}\left(\psi,^2_R-\psi,^2_T\right)=0\\
\textrm{G}_{33}& =2\left(\psi,_{TT}-\psi,_{RR}-\frac{1}{R}\,\psi,_{R}\right)+\textrm{G}_{22}=0.
\end{align}
We recommend the reader to distinguish the Roman R, the Ricci scalar, from
the italic $R$, one of the Einstein-Rosen coordinates.\\   
\indent Straight away one remarks that the component $\textrm{G}_{00}$ is identical to
$\textrm{G}_{11}$. Obviously only one of them is indispensable to obtain the
complete set of the field equations, say $\textrm{G}_{00}$. Just this equation yields one
of the differential equations determining $\psi$ and $\Gamma$, namely (\ref{enerdens}). Then of course we set in $\textrm{G}_{33}$ the component $\textrm{G}_{22}$ equal
zero, so to get the Bessel differential equation (\ref{sfield}). From the component
$\textrm{G}_{01}$ the last Einstein equation (\ref{currdens}) is read out.
The result is a compact set of the Einstein field equations:
\begin{align}
\label{sfield}
&\psi,_{TT}-\psi,_{RR}-\frac{1}{R}\,\psi,_{R}=0\\
\label{enerdens}
&\Gamma,_R=\tfrac{1}{2}\,R\,\left(\psi,^2_R+\psi,^2_T\right)\\
\label{currdens}
&\Gamma,_T=R\;\psi,_T\,\psi,_R
\end{align}
This set of equations has a familiar structure:
\begin{itemize}
\item Equation (\ref{sfield}) looks exactly like the usual \textit{wave
      equation} for the cylindrically symmetric \textit{massless scalar field}
      $\psi(T,R)$ propagating on a Minkowskian spacetime background.
\item The succeeding equation (\ref{enerdens}) describes the \textit{energy
      density} of the massless scalar field $\psi$ in the cylindrical
      coordinates.
\item At last, through equation (\ref{currdens}) the \textit{radial energy
      current density} is determined.
\end{itemize}
As explained in the beginning of this chapter, the foliated
(3+1)-dimensional general relativity has been reduced to a (1+1)-dimensional
model by the cylindrical symmetry. Further on, the model really is coupled to
a massless scalar field, since such a field solves the Einstein equations
written for the simplified cylindrical metric, namely the Einstein-Rosen
metric.\\
\indent Last but not least we wish to notice that the set of the Einstein equations,
reduced by the symmetry property of the Einstein-Rosen metric, can be deduced
in a easier and faster manner. The whole calculation can be slashed to the
task of complying 
\begin{equation}
\label{ricciten}
\textrm{R}_{\alpha\beta}=0.  
\end{equation}
The proof that the equations (\ref{einsteineqs}) are equivalent to the vanishing
Ricci tensor is very short. The main step is to multiply (\ref{einsteineqs})
from the left with the Einstein-Rosen metric $g^{ER}_{\alpha\beta}$. One gets
$\textrm{G}=-\textrm{R}$. This result is inserted in (\ref{einsteineqs})
wherefrom the following is given:
\begin{equation}
  \label{RG}
-\textrm{G}_{\alpha\beta}=\textrm{R}_{\alpha\beta}.
\end{equation}
In our case of a vacuum spacetime the assertion $\textrm{R}_{\alpha\beta}=0$
follows trivially, as $\textrm{G}_{\alpha\beta}=0$.\\
\indent Again the performed Ricci tensor for the Einstein-Rosen metric in
Appendix C is taken for eliciting the set of the Einstein equations. From
$\textrm{R}_{22}=0$ as also from $\textrm{R}_{33}=0$ we get the wave equation
for the massless scalar field (\ref{sfield}), both equation are
equivalent. Adding $\textrm{R}_{00}$ and $\textrm{R}_{11}$ together and
setting the sum equal to zero, (\ref{enerdens}) is obtained. Finally,
$\textrm{R}_{01}=\textrm{R}_{10}=0$ yields the energy current density (\ref{currdens}) of the
scalar field. The main simplification in computing the Einstein equations with the second
method is of course the fact that no Ricci scalar is needed.

%%% Local Variables: 
%%% mode: latex
%%% TeX-master: "liz"
%%% End: 

\clearemptydoublepage
%Zweites Kapitel: Loesung des Skalarfeldes, bisfield.tex

\chapter{Einstein-Rosen Wave}
\section{Solution of the Wave Equation}
In the present chapter we calculate the parameters $\psi(T,R)$ and
$\Gamma(T,R)$, which occur in the Einstein-Rosen metric and so determine the
Einstein-Rosen wave. The cylindrically symmetric wave equation (\ref{sfield})
is solved by a band of solutions. The parameter $\psi(T,R)$ is a
superposition of all allowed solutions. To keep
track of, we note the fundamental wave equation again:
\begin{equation} 
\label{ssfield} 
\psi,^{\phantom{}}_{TT}-\psi,^{\phantom{}}_{RR}-R^{\,-1}\;\psi,^{\phantom{}}_{R}=0
\end{equation}
The equation is obviously linear. The general solution of this differential
equation is the cylindrically symmetric massless scalar field $\psi(T,R)$
propagating on a Minkowskian background. The function $\psi(T,R)$ is the
sum over all modes. A mode is an oscillation with a particular frequency
$\omega=k=p, (c=\hbar=1)$, $k$ the wave number and $p$ the momentum of the
mode. As the frequency is a continuous parameter $(k\in\Ro)$ the sum
corresponds to an integration over all modes, i.e. over all $k\in\Ro$, or
equivalently over all momenta, $p\in\Ro$. Thus, the ansatz is
the well-known Fourier decomposition by frequencies. The method also is known
under the name ``separation of variables''. Thus we arrange the following
ansatz:
\begin{equation}
\label{ansatz}
\psi^{\phantom{}}_k(T,R)=\varphi^{\phantom{}}_k(T)\;\chi^{\phantom{}}_k(R)
\end{equation}
This ansatz is put in the differential equation (\ref{ssfield}) and thereupon
we divide by the product $\varphi^{\phantom{}}_k(T)\;\chi^{\phantom{}}_k(R)$,
in order to obtain
\begin{equation}
\label{diffansatz}
\frac{\varphi^{\phantom{}}_{k,\,TT}}{\varphi^{\phantom{}}_k}=\frac{\chi^{\phantom{}}_{k,\,RR}}{\chi^{\phantom{}}_k}+\frac{1}{R}\,
\frac{\chi^{\phantom{}}_{k,\,R}}{\chi^{\phantom{}}_k}\;.
\end{equation}
The ansatz (\ref{ansatz}) presumes the functions
$\varphi^{\phantom{}}_k$, $\chi^{\phantom{}}_k$ to depend only on $T$ and
$R$ respectively. This fact constrains the right and the left side of the equation
(\ref{diffansatz}) to be equal to a constant $k^2$ with $k\in\Ro$. Otherwise
the treated differential expression does not hold anymore. Then we set
\begin{align}
\label{timediff}
&\frac{\varphi^{\phantom{}}_{k,\,TT}}{\varphi^{\phantom{}}_k}=-k^2\\
\label{spacediff}
&\frac{\chi^{\phantom{}}_{k,\,RR}}{\chi^{\phantom{}}_k}+\frac{1}{R}\,
\frac{\chi^{\phantom{}}_{k,\,R}}{\chi^{\phantom{}}_k}=-k^2,\qquad k\in\Ro.
\end{align}
The first relation is the ordinary free harmonic oscillator differential equation
with oscillation frequency $k$. This equation is solved by a linear combination
of both solutions
\begin{eqnarray}
\nn \varphi^{(1)}_k(T)&=&\textrm{A}(k)\,\ex^{ikT}\\
    \varphi^{(2)}_k(T)&=&\textrm{B}(k)\,\ex^{-ikT}.\label{phisols}
\end{eqnarray} 
So we find for the general solution of the time part of (\ref{ssfield}) the
announced overlay:
\begin{equation} 
\label{timegensol}
\varphi^{\phantom{}}_k(T)=\textrm{A}(k)\,\ex^{ikT}+\textrm{B}(k)\,\ex^{-ikT}.
\end{equation}
The functions $\textrm{A}(k)$ and $\textrm{B}(k)$ are in relationship with each other. As we
know, gravitational waves, like electromagnetic waves, are real valued waves. From
this it follows that the solution of the differential equation (\ref{ssfield})
has to be real. The function $\psi(T,R)$ is real if and only if the following holds:
\begin{equation}
\textrm{B}(k)=\textrm{A}^{\ast}(k),
\end{equation}
whereas $\textrm{A}^{\ast}(k)$ is the complex conjugate of $\textrm{A}(k)$.
Thereupon the solution to the time part of the wave equation reads:
\begin{equation}
\varphi^{\phantom{}}_k(T)=\textrm{A}(k)\,\ex^{ikT}+\textrm{A}^{\ast}(k)\,\ex^{-ikT}.
\end{equation}
\indent Subsequently we consecrate ourselves to the spatial part
(\ref{spacediff}) of the wave equation. Rewriting the equation into the
form
\begin{equation} 
\label{bspacediff}
R^2\;\chi^{\phantom{}}_{k,\,RR}+R\;\chi^{\phantom{}}_{k,\,R}+k^2\,R^2\;\chi^{\phantom{}}_k=0\quad,
\end{equation}
we notice the evident similarity to the Bessel differential equation
\begin{equation}
\label{besseleq}
z^2\;\frac{d^2}{dz^2}\;w(z)+z\;\frac{d}{dz}\;w(z)+(z^2-n^2)\;w(z)=0.
\end{equation}    
In addition we set in (\ref{besseleq}) $n=0$ and in (\ref{bspacediff})
$s\doteq kR,\ s\in\R$.
The spatial differential equation is indeed the Bessel differential equation
with index $n=0$.
\begin{equation}
\label{bsdiff}
s^2\;\frac{d^2}{dR^2}\;\chi^{\phantom{}}_k+s\;\frac{d}{dR}\;\chi^{\phantom{}}_k+s^2\;\chi^{\phantom{}}_k=0,
\end{equation}
with $s=s(k;R)$. The Bessel function of first order solve the equations with
$n=0$. These are (see\cite{abramo72}): $\textrm{J}_0(kR)$,
$\textrm{Y}_0(kR)$, $\textrm{H}^{(+)}_0(kR)$, $\textrm{H}^{(-)}_0(kR)$, where
$\textrm{H}^{(\pm)}_0(kR)=\textrm{J}_0(kR)\pm i\textrm{Y}_0(kR)$.
For the solution $\psi(T,R)$ the following boundary
condition is given \cite{kuchar71}:
\begin{equation}
\label{boundcond}
\lim_{R\to 0}\;R\;\psi^{\phantom{}},_{T}\;\psi^{\phantom{}},_{R}=0
\end{equation}
Since the Weber function $\textrm{Y}_0(kR)$ contains the logarithm function
depending just on $R$, all expressions involving the considered function are
invalid solutions, as the limes in (\ref{boundcond}) does not exist. For this
reason the cylindrically symmetric Bessel function $\textrm{J}_0(kR)$ is the
last possibility to solve the spatial differential equation. For the further
considerations we choose the representation for the Bessel function
$\textrm{J}_0(kR)$ given in appendix A, namely (\ref{besseln}), but with
$z\doteq kR$ and $n=0$. Setting $\sin\varphi=-\cos\vartheta$ with
$\vartheta\doteq\varphi+\frac{\pi}{2}$ and applying what is proved in
(\ref{proofcos}), we get for $\textrm{J}_0(kR)$: 
\begin{equation}
\label{coskRbessel}
\textrm{J}_0(kR)=\frac{1}{2\pi}\int^{2\pi}_0d\varphi\;\ex^{\,ikR\cos\varphi}
\end{equation}
Inserting $\textrm{J}_0(kR)$ in (\ref{boundcond}) we obtain 
\begin{equation}
\label{calcond}
\lim_{R\to 0}\;R\;\varphi^{\phantom{}}_{k,\,T}(T)\;\textrm{J}_0(kR)\;\varphi^{\phantom{}}_k(T)\;\frac{\partial}{\partial R}\;\textrm{J}_0(kR)=0.
\end{equation}
Thus, the Bessel function $\textrm{J}_0(kR)$ solves the spatial differential
equation and the resulting wave function $\psi^{\phantom{}}_k(T,R)$ the
boundary condition too.
\begin{equation}
\label{mode1}
\psi_k(T,R)=\textrm{J}_0(kR)\;\left[\textrm{A}(k)\ \ex^{ikT}+\textrm{A}^{\ast}(k) \ex^{-ikT}\right].  
\end{equation}
The above function is the solution for a specific mode, namely the one with
frequency $\omega=k$. To obtain the general solution for the wave equation
(\ref{ssfield}), we have to sum together all modes, thus to integrate over $k$,
accordingly to the Fourier theorem. Therewith the general solution of the
wave equation for the cylindrically symmetric massless scalar field
propagating on a Minkowskian spacetime is: 
\begin{equation}
\label{gensolu}
\psi(T,R)=\int^\infty_0dk\;\mjo\left[\textrm{A}(k)\,\ex^{ikT}+\textrm{A}^{\ast}(k)\,\ex^{-ikT}\right].
\end{equation}
If we compare the solution (\ref{gensolu}) with the Hankel transformation
(\ref{hankel}) in appendix A for the case $n=0$, we remark a close similitude. In fact the function
(\ref{gensolu}) can be written as the Hankel transformation of
$\frac{1}{k}\varphi_k(T)$, that is to say
\begin{equation}
\label{hankelsol}
\psi(T,R)=\int^\infty_0dk\;k\;\mjo\left[\frac{\textrm{A}(k)}{k}\,\ex^{ikT}+\frac{\textrm{A}^{\ast}(k)}{k}\,\ex^{-ikT}\right].
\end{equation}
Because of the close relation of $\psi(T,R)$ to the Hankel transformation, we
define newly the expression $\frac{1}{k}\varphi_k(T)$ . We set
\begin{equation}
  \label{trafo}
\tilde{\psi}(k,T)\doteq\tfrac{1}{k}\varphi_k(T) 
\end{equation}
with the Hankel coefficients
\begin{align}
  \label{afield}
a(k)\doteq\frac{\textrm{A}(k)}{k},\qquad a^{\ast}(k)\doteq\frac{\textrm{A}^{\ast}(k)}{k}.  
\end{align}
Therewith the scalar field can be written as the Hankel transformation of $\tilde{\psi}(k,T)$:
\begin{eqnarray}
\label{hankelgensol}
\nn \psi(T,R)&=&\int^\infty_0dk\;k\;\mjo\;\tilde{\psi}(k,T)\\
             &=&\int^\infty_0dk\;k\;\mjo\left[a(k)\,\ex^{ikT}+a^{\ast}(k)\,\ex^{-ikT}\right].
\end{eqnarray}
In order to be sure that the Hankel transformation of $\psi(T,R)$ is a well defined
function, it is essential to give a look to the property the Hankel
coefficients $a(k)$ and $a^{\ast}(k)$ have to fulfill. As we can learn from appendix A, the two dimensional
Fourier transformation with cylindrical symmetry is just the Hankel transformation -
we can imagine that the Hankel transformation of $\psi(T,R)$ exists only if
$\tilde{\psi}(k,T)$, and in detail $a(k)=\frac{\textrm{A}(k)}{k}$ and
$a^{\ast}(k)=\frac{\textrm{A}^{\ast}(k)}{k}$,
is integrable. In addition to this requirement, we would like that the
function we get through the transformation, namely $\psi(T,R)$, is again
integrable over the radius $R$. This would assure the possibility to
transform back $\psi(T,R)$ to the functions $a(k)$ and
$a^{\ast}(k)$. This means mathematically expressed:
\begin{align}
  \label{fourinv}
&\mathcal{H}:\psi(R;T)\rightarrow\tilde{\psi}(k;T) &(\textrm{Hankel transformation})\\
&\mathcal{H}^{-1}:\tilde{\psi}(T;k)\rightarrow\psi(R;T) &(\textrm{Inverse transformation})  
\end{align}
and thus the identity
\begin{equation}
\label{id}
\mathcal{H}^{-1}\left[\mathcal{H}\left[\psi(R;T)\right]\right]=\mathcal{H}^{-1}\left[\tilde{\psi}(T;k)\right]=\psi(T,R).
\end{equation}
In case of the Fourier transformation there is a function space $\mathcal{S}$ (Schwartz space), which guarantees the existence of the
inverse map of a Fourier transformed function, if the treated function is
element of $\mathcal{S}$. This is ensured by the following theorem:\\
\\
\textbf{Fourier inversion theorem}:\\
The Fourier transform is a bicontinuous bijection from $\mathcal{S}(\R^n)$ onto
$\mathcal{S}(\R^n)$. Its inverse map is the inverse Fourier transform,
i.e. (\v{f})\^{} = f = (\^{f})\v{}.
The Fourier transform and the inverse map are denoted by  \^{f} and \v{f}
respectively \cite{reed75}.\\
\\ 
Since the Hankel transformation is nothing but the two dimensional Fourier
transformation with cylindrical symmetry it is assumed that an analogous
theorem holds, which presupposes the Hankel coefficients to be element of the
Schwartz space in order to assure the identity (\ref{id}).  
So, if it is set $\textrm{A}(k)$ and $\textrm{A}^{\ast}(k)$ to be
element of the Schwartz space, then the functions $a(k)$ and
$a^{\ast}(k)$ are too, since the functions in $\mathcal{S}$ fall off faster
than every power of their argument increasing to infinity. So we managed to
assure the integrability of the wave function $\psi(T,R)$ and, of course, of
its Hankel transformed function $\tilde{\psi}(k,T)$ also. 
\section{Finiteness of the Energy $\Gamma(T,R)$}
The integrability property of the wave function $\psi(T,R)$ is a vital
condition for the function $\Gamma(T,R)$. As it is shown in the appendix B, it
turns out that the energy of the scalar field, contained in a disk $\Delta
z=1$, is given (up to a factor $2\pi$) by this same function 
\begin{equation}
  \label{gamma}
\Gamma(T,R)=\int^R_0dR'\ \Gamma,_{R'}(T,R').  
\end{equation}
It is now important to prove that the energy remains finite for the radius
$R$ at infinity\footnote{The next equation is given by the expression
  (\ref{enerdens}) for the energy density of the massless scalar field $\psi(T,R)$.}.
\begin{equation}
  \label{gammainfty}
\Gamma(T,\infty)=\tfrac{1}{2}\int^\infty_0dR\;R\,\left(\psi,^2_T+\psi,^2_R\right).  
\end{equation}
Obviously, the properties of $\psi(T,R)$ play a main role for the finiteness of
the energy.
We first start with treating a single mode of $\psi(T,R)$ with the
specific frequency $k$.
\begin{equation}
\label{mode2}
\psi_k(T,R)=\textrm{J}_0(kR)\;\left[\textrm{A}(k)\ \ex^{ikT}+\textrm{A}^{\ast}(k) \ex^{-ikT}\right]. 
\end{equation}
As we will integrate over $R$, we concentrate for the time being only on the
$R$-depending part 
\begin{equation}
\label{Rpart}
\psi_k(R)=\textrm{J}_0(kR). 
\end{equation}
The plain wave is not normalizable as is generally known\footnote{For eq. (\ref{plainwave}): The orthonormality relation is presented in appendix A. Replace $x$ with $R$ in equation (\ref{hankortho}).}. 
\begin{equation}
\label{plainwave}
||\psi_k(R)||^2=\int^\infty_0dR\;R\;\textrm{J}_0(kR)\textrm{J}_0(k'R)=\frac{1}{k'}\delta(k-k').
\end{equation}
In such cases a wave packet is introduced, which, for instant, could be
formed by a Gauss profile  $\textrm{A}(k)$. This function is an element of the Schwartz space, and so falls off
faster than every power of the argument $k$.
\begin{equation}
\label{wavepacket}
\bar{\psi}(R)=\int^\infty_0dk\;\textrm{A}(k)\textrm{J}_0(kR). 
\end{equation}
Now the wave is normalizable, since the following integral is smaller than
infinity due to the wave packet, which vanishes at infinity.
\begin{equation}
\label{norm1}
||\bar{\psi}(R)||^2=\int^\infty_0dk\;|\textrm{A}(k)|^2<\infty.  
\end{equation}
Comparing (\ref{wavepacket}) with (\ref{mode2}) or directly with (\ref{gensolu})
we remark promptly that the wave function (\ref{gensolu}) is furnished already
with its wave packet, and is therefore normalizable:
\begin{eqnarray}
  \label{finitnorm}
||\psi(T,R)||^2&=&\int^\infty_0dR\;R\;\psi^2(T,R)\\
&=&\int^\infty_0dk\;k\;\left[a(k)\ \ex^{ikT}+a^{\ast}(k)\ \ex^{-ikT}\right]^2.  
\end{eqnarray}
So we can verify the finiteness of the energy $\Gamma(T,R)$ if we integrate
over the whole space at the fixed time $T$. For the derivatives of the wave
function with respect to $T$ and $R$ we get
\begin{eqnarray}
  \label{derivT}
\psi,_T&=&i\int^\infty_0dk\;k^2\;\textrm{J}_0(kR)\,\varphi_{(-)}(k;T)\\
  \label{derivR} 
\psi,_R&=&-\int^\infty_0dk\;k^2\;\textrm{J}_1(kR)\,\varphi_{(+)}(k;T),   
\end{eqnarray}
where we define 
\begin{equation}
  \label{phi}
\varphi^{(\mp)}\doteq a(k)\ \ex^{ikT}\mp a^{\ast}(k)\ \ex^{-ikT}.  
\end{equation}
If we square both derivatives and integrate them over the
radius $R$, we obtain: 
\begin{eqnarray}
\label{squdev}
\int^\infty_0dR\;R\;\psi^2,_T(T,R)&=&-\int^\infty_0dk\;k^3\,\varphi_{(-)}^2(k;T)\\
\int^\infty_0dR\;R\;\psi^2,_R(T,R)&=&\int^\infty_0dk\;k^3\,\varphi_{(+)}^2(k;T), 
\end{eqnarray}
where we have used the orthonormality relation for the Bessel
function\footnote{See (\ref{hankortho}), for the case $x=R$.}
\begin{equation}
\label{ortho}
\int^\infty_0dR\;R\;\textrm{J}_n(kR)\;\textrm{J}_n(k'R)=\frac{1}{k'}\,\delta(k-k'),
\qquad\forall n\in\mathbb{Z}. 
\end{equation}
Both integrals are put now in (\ref{gammainfty}), so to get the finite quantity
\begin{eqnarray}
\label{inftyenergy}
\nn \Gamma(T,\infty)&=&2\int^\infty_0dk\;k^3\;a(k)\;a^{\ast}(k)\\
\nn &=&2\int^\infty_0dk\;k^3\;|a(k)|^2\\
&=&2\int^\infty_0dk\;k\;|\textrm{A}(k)|^2,\qquad \textrm{A}(k), a(k)\in\mathcal{S}.  
\end{eqnarray}
\rightline{QED}
We conclude the present chapter affirming that the elicited wave function
$\psi(T,R)$ is well-defined, since for the two cases, $R\to 0$ and
$R\to\infty$, the function remains bounded providing at the same time its
normalizibility and a finite energy of the field.

%%% Local Variables: 
%%% mode: latex
%%% TeX-master: "liz"
%%% End: 

\clearemptydoublepage
%ADM formalism, ADM.tex

\chapter{ADM Formalism}
In the previous chapters we have presented how the Einstein equations can be
solved if the symmetries of the metric are known. In our case it is about the Einstein-Rosen
spacetime and so the Einstein equations have been solved for the
Einstein-Rosen metric (\ref{ermetric}). The solution to the differential equations
of second order is then, as we have seen, the massless scalar
field $\psi$. In general the Einstein equations 
\begin{equation}
  \label{equES}
\textrm{R}_{\alpha\beta}-\tfrac{1}{2}g_{\alpha\beta}\textrm{R}=0
\end{equation}   
are obtained by varying the following action with respect to the metric
$\g_{\alpha\beta}$, and discarding the resulting divergences\footnote{See \cite{weinberg72}.}:
\begin{equation}
  \label{action}
  S[g_{\alpha\beta}(x)]=\frac{1}{16\pi G}\int d^4x\;\mathcal{L}=\frac{1}{16\pi G}\int\sqrt{g(x)}\,\textrm{R}(x)\,d^4x,
\end{equation}
where $\mathcal{L}$ stands for the Lagrangian density and $g(x)$ for the
determinant of the metric $\g_{\alpha\beta}$ at the
spacetime point $x$. In the Einstein equations above,
$\textrm{R}_{\alpha\beta}$ denotes the Ricci tensor, $\textrm{R}$ the Ricci
scalar and finally $\textrm{G}$ the universal gravitational constant. In the early sixties of
the last century R. Arnowitt, S. Deser and C.W. Misner, short ADM, developed
the so called canonical formalism for general relativity \cite{ADM62}. A
reason for constructing such a formalism to describe dynamic in general
relativity is that an action containing canonical conjugate variables is a possible basis
to develop a quantum theory of gravity by canonical methods.\\
\indent In order to introduce canonical variables, a frame is needed, where they can be
related on. To this end, spacetime, in the ADM formalism, is split into three
spatial dimensions and one time dimension, in literature shortly noted by
(3+1). One can now attempt to imagine spacetime as a composite of
hypersurfaces (3-space surfaces), which are fixed at arbitrary points in
time. So a hypersurface $\mathcal{G}_{t_0}$  could be fixed at the time
$t=t_0$, just as the hypersurfaces $\mathcal{G}_{t_k}$ at an arbitrary time
$t=t_k$. Here we would like to consider two specific hypersurfaces
$\mathcal{G}_{t_0}$ and $\mathcal{G}_{t_1}$ to different times
$t_0<t_1$. In the following we try to give an idea how points in the first
hypersurface $\mathcal{G}_{t_0}$ are connected with the corresponding points
in the later hypersurface $\mathcal{G}_{t_1}$. In the ADM
formalism two functions  are brought in, which measure the translation in
spacetime. The lapse of the proper time between the 'lower'
$\mathcal{G}_{t_0}$ and the 'upper' $\mathcal{G}_{t_1}$ hypersurface is
measured by the so called lapse function $\textrm{N}(t,x_1,x_2,x_3)$. On the
other hand the shift function $\textrm{N}^i(t,x_1,x_2,x_3)$ measures the
spatial shift, i.e. this function specifies the position in the 'upper'
hypersurface whereto the point $x^i$ of the 'lower' hypersurface have to be
placed. The infinitesimal translation in spacetime is given by the interval
$ds^2$ between the point $x^{\alpha}=(t,x^i)$ and the point
$x^{\beta}+dx^{\beta}=(t+dt,x^i+dx^i)$ in the 'upper' hypersurface, namely by
the relation
\begin{equation}
  \label{intervall}
ds^2=g_{ik}(dx^i+\textrm{N}^idt)(dx^k+\textrm{N}^kdt)-(\textrm{N}dt)^2=g_{\alpha\beta}dx^{\alpha}dx^{\beta}.  
\end{equation}
Thus, the 4-metric has been constructed out of the 3-metric $^3g_{ik}$, the
lapse function and the shift function:
\begin{equation}
  \label{lsmetric}
\left(\begin{array}{cc}
g_{00} & g_{0k} \\
g_{i0} & ^3g_{ik} \\
\end{array}\right)
=
\left(\begin{array}{cc}
\textrm{N}_s\textrm{N}^s-\textrm{N}^2 & \textrm{N}_{k} \\
\textrm{N}_{i} & ^3g_{ik} \\
\end{array}\right).
\end{equation}
If total divergences are discarded from the Lagrangian density of the action
(\ref{action}), then the functional $\textrm{S}$ can be brought into the form
\begin{equation}
  \label{admaction}
\textrm{S}=\int dt\int d^3x\,\textrm{N}g^{1/2}(\textrm{K}_{ik}\textrm{K}^{ik}-\textrm{K}^2+\textrm{R}).  
\end{equation}
Here, $\textrm{R}$ is again the Ricci scalar and $g^{1/2}$ the square root of
the determinant of the 3-metric $g_{ik}$. $\textrm{K}^2$ is the squared trace
of the tensor $\textrm{K}_{ik}$:
$\textrm{K}=\textrm{K}^i_i=g^{ik}\textrm{K}_{ik}$. The new second order tensor
$\textrm{K}_{ik}$ is the extrinsic curvature of the spacelike hypersurface. The
extrinsic curvature is calculated out of the 3-metric, the lapse and the
shift function, according to the formula
\begin{equation}
  \label{curv}
\textrm{K}_{ik}=\frac{1}{2\textrm{N}}(\textrm{N}_{i|k}+\textrm{N}_{k|i}-g_{ik,0}),  
\end{equation}
where the stroke means the covariant derivative in 3-space. As an illustration
to the extrinsic curvature, we cite \cite{mtw73}:\\
\textit{The extrinsic curvature measures the fractional shrinkage and deformation of a
figure lying in the spacelike hypersurface $\Sigma$ that takes place when each
point is carried forward a unit interval of proper time ``normal'' to the
hypersurface out into the enveloping spacetime. No enveloping spacetime? No
extrinsic curvature!}\\
Varying the action (\ref{admaction}) with respect to $g_{ik,0}$ we obtain the
momentum $\pi^{ik}$, which is canonical conjugate to the 3-metric $g_{ik}$:
\begin{equation}
  \label{pi}
\pi^{ik}=\frac{\delta\mathcal{L}}{\delta g_{ik,0}}=-g^{1/2}(\textrm{K}^{ik}-\textrm{K}g^{ik}), 
\end{equation} 
where the Lagrange density $\mathcal{L}$ is the integrand of the action (\ref{admaction})
and $16\pi\textrm{G}=1$ is set. Therewith it is scheduled what, in the ADM
formalism, the generalized coordinates and the belonging canonical
momentum are, namely the 3-metric $g_{ik}$ and the momentum $\pi^{ik}$. Just the
introduction of the momentum
\begin{equation}
\label{lageq}
\pi^{ik}=\frac{\delta\mathcal{L}}{\delta g_{ik,0}}
\end{equation}
makes it possible to transform the second order partial differential
equations of the Einstein equations to an equivalent set of first order differential
equations. This set will then depend on the dynamic variables $g_{ik}$ and
$\pi^{ik}$. The action functional is then converted into the Hamiltonian form,
which is more convenient for the developing of a quantum theory of
gravity by canonical methods, since the Hamiltonian enters the action.
\begin{equation}
  \label{haction} 
\textrm{S}=\int dt\int
d^3x(\pi^{ik}g_{ik,0}-\textrm{N}\mathcal{H}-\textrm{N}_i\mathcal{H}^i),  
\end{equation}
with $\textrm{S}=\textrm{S}[g_{ik},\pi^{ik},\textrm{N},\textrm{N}_i]$. The
lapse $\textrm{N}$ and the shift function $\textrm{N}_i$ in (\ref{haction})
play the role of Lagrange multipliers, whereby the function $\mathcal{H}$ is
the so called super-Hamiltonian and the function $\mathcal{H}^i$ the
supermomentum. These quantities are expressed by the canonical conjugate
variables $g_{ik}$ and $\pi^{ik}$,
\begin{eqnarray}
  \label{superham}
\mathcal{H}&=&g^{-1/2}\left(\pi^{ik}\pi_{ik}-\tfrac{1}{2}\pi^2\right)-g^{1/2}\textrm{R},\\
  \label{supermoment}
\mathcal{H}^i&=&-2 \pi^{ik},_k-g^{il}\left(2g_{jl,k}-g_{jk,l}\right)\pi^{jk}.
\end{eqnarray}
Analogously to $\textrm{K}$ in (\ref{admaction}), $\pi$ denotes the trace of
$\pi^{ik}$ and so $\pi^2$ is nothing but the square of the trace of
$\pi^{ik}$: $\pi=\pi^i_i=g_{ik}\pi^{ik}$. The commas in the right hand side of
the equation of the supermomentum represent the common partial derivative with
respect to the space coordinates $k,l=1,2,3$. Now, if the action (\ref{haction})
is varied with respect to the Lagrange multipliers $\textrm{N}$ and
$\textrm{N}_i$, then the constraints of the system are obtained:
\begin{align}
  \label{constraintH}
&\mathcal{H}=0,\\
  \label{constraintHi}
&\mathcal{H}^i=0.  
\end{align}
Varying whereas with respect to the dynamical variables $g_{ik}$ and
$\pi^{ik}$, the set of canonical equations is yield, which is well known
from the Hamilton formalism and which, apart from the constraints, replace the
Einstein equations in the form (\ref{equES}).
\begin{equation}
  \label{Heq1}
g_{ik,0}=\frac{\delta\textrm{H}}{\delta\pi^{ik}},\qquad\pi^{ik},_0=\frac{\delta\textrm{H}}{\delta
g_{ik}},
\end{equation}
with the Hamiltonian
\begin{equation}
  \label{Hamiltonian}
\textrm{H}=\int d^3x\left(\textrm{N}\mathcal{H}+\textrm{N}_i\mathcal{H}^i\right)  
\end{equation}
We can see the relation (\ref{Hamiltonian}) by recalling the definition of the Hamilton function:
\begin{equation}
  \label{Hamfct}
\textrm{H}=p^i\dot{q}_i-\textrm{L},  
\end{equation}
where $p^i$ are the canonical conjugate momenta to the generalized coordinates $q_i$. In our case we have then,
\begin{eqnarray}
\label{L}
\textrm{L}&=&\int d^3x\left(\pi^{ik}g_{ik,0}-\textrm{N}\mathcal{H}-\textrm{N}_i\mathcal{H}^i\right),\\
\label{pq}
p^i\dot{q}_i&=&\int d^3x\;\pi^{ik}g_{ik,0},
\end{eqnarray}
which reproduce equation (\ref{Hamiltonian}) if inserted in (\ref{Hamfct}).
\section{Reduced ADM Action for the Cylindrically\\ Symmetric Spacetime}\label{secADMCSS}
We dedicate this section to the ADM action for the cylindrically
symmetric spacetime. Due to the symmetries of the model the 4-metric
simplifies drastically, as it is shown in chapter one, see (\ref{subsecconfdef}),
(\ref{subsecIS}). The first simplification is that the cylindrically symmetric
metric depends only on the coordinates $t$ and $r$, since it is invariant with
respect to changes in the coordinates $\varphi$ and $z$. Secondly, the metric
acquires diagonal form by specific conformal transformations \footnote{See
(\ref{diag}), (\ref{barmetric}).}, a fact which simplifies the task of solving the Einstein
equations. These properties of course have influence on all quantities of the
ADM  formalism. We start with the lapse and the shift functions, which are
related through equation (\ref{lsmetric}) to the 4-metric, in our case the
cylindrically symmetric one. The only component of the cylindrically symmetric
metric, which in one hand is non vanishing and in the other hand is in
relation with the lapse and the shift functions, is (\ref{barmetric})
\begin{equation}
  \label{g00}
g_{00}=-\Omega^2(t,r),  
\end{equation}
where in the following we use the coordinates $t$ and $r$ in the conformal
transformed metric instead of $\bar{t}$ and $\bar{r}$.
This component is only dependent on the coordinate $t$ and $r$, such that also
the Lagrangian multipliers $\textrm{N}(t,r)$ and $\textrm{N}^i(t,r)$ of the
ADM action in the Hamiltonian form only depend on these
coordinates. Further on two components of the shift function drop out due to
the symmetries, namely the ones which measures the changes in the $\varphi$
and $z$ coordinates. So, only the lapse function $\textrm{N}(t,r)$ and one
shift function $\textrm{N}^1(t,r)$, namely that one which measures
translations in the radius direction, are left. Also the 3-metric depends 
only on the time and radius coordinates and as part of the conformal
transformed metric, it is in diagonal form, see (\ref{barmetric}). Since the canonical
conjugate momenta are related to the 3-metric by the equation
\begin{equation}
  \label{momeq}
\pi^{ik}=\frac{\delta\mathcal{L}}{\delta g_{ik,0}},  
\end{equation}
one can imagine that the momenta inherit the properties of the 3-metric. In
\cite{kuchar71} the symmetry reduction of the momenta is presented at
length, so that we note only the results. The non vanishing components of the
momentum tensor $\pi^{ik}$ are the following, with $1,2,3$ labeling $r,\varphi,z$:
\begin{equation}
  \label{nonvanmom}
\pi^{11}(t,r),\qquad\pi^{22}(t,r),\qquad\pi^{33}(t,r).
\end{equation}
As expected, there remain only the components left, which are canonical
conjugate to the corresponding non vanishing components of the 3-metric,
\begin{equation}
  \label{nonvanmet}
\g_{11}(t,r),\qquad\g_{22}(t,r),\qquad\g_{33}(t,r).
\end{equation} 
If in the following discussion we take the metric
(\ref{barmetrik}) suggested in \cite{kuchar71} as an example for a
cylindrically symmetric spacetime, we see that the metric is given by the quantities $\gamma$, $R$ and $\psi$. The
equations of motion, i.e the Einstein equations in the ADM form will finally be
differential equations of those variables. So it is throughout correct to
look at these quantities as generalized coordinates  and to note the canonical
conjugate momenta with respect to those generalized coordinates
\begin{align}
  \label{pgrmom}
\pi^{11}(t,r)& =\pi_{\gamma}\,\ex^{\psi-\gamma},& \pi^{22}(t,r)&
=\tfrac{1}{2}R^{-1}\pi_R\,\ex^{\psi},& \pi^{33}(t,r)&
=(\pi_{\gamma}+\tfrac{1}{2}R\,\pi_R+\pi_{\psi})\ex^{-\psi},  
\end{align}
where our calculation yields for the component $\pi^{22}$ a factor $R^{-1}$
instead of the factor $R$ noted in the paper \cite{kuchar71}. Thus, it has
been achieved to bring the expression $\pi^{ik}g_{ik,0}$ of the action
(\ref{haction}) in the following canonical form:
\begin{equation}
  \label{canpig}
\pi^{ik}g_{ik,0}=\pi_{\gamma}\dot{\gamma}+\pi_R\dot{R}+\pi_{\psi}\dot{\psi}, 
\end{equation}
where this form holds exclusively for the cylindrically symmetric
spacetime and especially for the metric (\ref{barmetrik}) depending on
$\gamma$, $R$ and $\psi$. 
The dots in equation (\ref{canpig}) denote the partial derivative
$\partial_t$. If (\ref{pgrmom}) is solved for the momenta conjugate to the
generalized coordinates $\gamma$, $R$ and $\psi$ then we get
\begin{eqnarray}
  \label{pigamma}
\pi_{\gamma}&=&\ex^{\psi-\gamma}\pi^{11},\\
  \label{piR}
\pi_R&=&2R\,\ex^{-\psi}\pi^{22},\\
  \label{pipsi}
\pi_{\psi}&=&\pi^{33}\,\ex^{\psi}-\pi^{22}R^2\,\ex^{-\psi}-\pi^{11}\,\ex^{\gamma-\psi}.  
\end{eqnarray}
Now it is straight forward to calculate the super-Hamiltonian and the
supermomentum. To this end one inserts in (\ref{superham}) for $\pi^{ik}$ the three
non vanishing momenta $\pi^{11}$, $\pi^{22}$ and $\pi^{33}$ and for $g_{ik}$ the
non vanishing components of the 3-metric,
\begin{equation}
  \label{nonvcomp}
g_{11}=\ex^{\gamma-\psi},\qquad g_{22}=R^2\ex^{-\psi}, \qquad g_{33}=\ex^{\psi}.   
\end{equation}
The same is inserted in (\ref{supermoment}) for the computation of the
supermomenta. To simplify the reduced ADM action anymore, the super-Hamilton
and the lapse function as also the supermomentum and the shift function are
rescaled respectively with the factors $\ex^{1/2(\gamma-\psi)}$ and
$\ex^{\psi-\gamma}$. The rescaled constraints read
\begin{eqnarray}
  \label{Htilde}
\tilde{\mathcal{H}}&=&-\pi_{\gamma}\pi_{R}+\tfrac{1}{2}R^{-1}\pi_{\psi}^2+2R''-\gamma'R'+\tfrac{1}{2}R\psi'{^2},\\
  \label{H1tilde}
\tilde{\mathcal{H}_1}&=&-2\pi_{\gamma}'+\gamma'\pi_{\gamma}+R'\pi_{R}+\psi'\pi_{\gamma},\\
  \label{H23tilde} 
\tilde{\mathcal{H}_2}&=&\tilde{\mathcal{H}_3}=0,  
\end{eqnarray}
where the primes stand for the derivative with respect to the radius
coordinate $r$.
It is not surprising to see only one supermomentum, which does not
vanish. We should take into account, there is only one non vanishing shift function,
namely $\textrm{N}_1(t,r)$. The variation of the ADM action in Hamiltonian
form with respect to the rescaled Lagrange multipliers yields only the
following two constraints,
\begin{eqnarray}
  \label{tildecons}
\tilde{\mathcal{H}}&=&0,\\
  \label{tildecons1}
\tilde{\mathcal{H}}_1&=&0.  
\end{eqnarray}
Inserting (\ref{canpig}) and the rescaled constraints in the expression
(\ref{haction}) for the ADM action, we get finally the symmetric reduced ADM
action for the case of the cylindrically symmetric spacetime described by the
metric (\ref{barmetrik}).
\begin{equation}
  \label{redaction}
S=2\pi\int^{\infty}_{-\infty}dt\int^{\infty}_{0}dr\,(\pi_{\gamma}\dot{\gamma}+\pi_R\dot{R}+\pi_{\psi}\dot{\psi}-\tilde{\textrm{N}}\tilde{\mathcal{H}}-\tilde{\textrm{N}}^1\tilde{\mathcal{H}_1}).  
\end{equation}
\section{Reduced ADM Action for the  Einstein-Rosen\\ Wave}
After bringing the general cylindrically symmetric metric (\ref{allgmetric})
into diagonal form by a conformal and a coordinate transformation into the
inertial system, a further simplification of the diagonal metric was
performed. By another conformal transformation of the coordinates\footnote{See
section (\ref{secERM}) and especially (\ref{subsecCCT}).}, the Einstein-Rosen
coordinates were introduced. Exactly this step has to be repeated, i.e. to
introduce the Einstein-Rosen coordinates in the canonical formalism such that
the structure of the super-Hamilton and the supermomentum get more
simplified. The Einstein-Rosen radius $R$ is already present in the canonical
formalism, so that it remains only to enter the Einstein-Rosen time $T$. To
this end the function $T$ is defined by the extrinsic curvature\footnote{See
  section VII in \cite{kuchar71}} and is converted to a canonical coordinate
by a canonical transformation. The canonical coordinate $T$ is then identified
with the momentum $\pi_{\gamma}$, according to
\begin{equation}
  \label{Tpi}
\pi_{\gamma}=-T',  
\end{equation}
where the prime denotes the derivative with respect to $r$. The canonical
transformation also allows to identify $-\gamma'$ in (\ref{Htilde}) and (\ref{H1tilde})
with the canonically conjugate momentum of the Einstein-Rosen time,
\begin{equation}
  \label{piTg}
\pi_T=-\gamma'.  
\end{equation}
In section \ref{subsecCCT} we have seen that the introduction of the
Einstein-Rosen coordinates changed $\gamma$ to the 'energy' function $\Gamma$
of the massless scalar field following the relation (\ref{gG}). Through this
transformation also the momentum $\pi_T$ changes according to (\ref{piTg}). The result
is the introduction of a new momentum $\Pi_{T}$, which forces the
implementation of a new momentum canonically conjugate to $R$, namely
$\Pi_R$. The relations between $\pi_T$ and $\Pi_{T}$ and between $\pi_R$
and $\Pi_R$ are taken from \cite{kuchar71}:
\begin{eqnarray}
  \label{PiT} 
\Pi_T&=&\pi_T+[\ln(R'{^2}+T'{^2})]',\\
  \label{PiR}
\Pi_R&=&\pi_R+\left[\ln\left(\frac{R'+T'}{R'-T'}\right)\right]'.
\end{eqnarray}
So the new set of the canonical variables for the reduced action is:
$(T,\Pi_T)$, $(R,\Pi_R)$, $(\psi,\pi_{\psi})$. The canonical momentum
$\pi_{\psi}$ is unaffected by the canonical transformation, since $\psi$
is untouched by it. Analogously $\psi$ did not change by the conformal
transformation (\ref{conftrans}).\\
\indent Introducing the new canonical variables $(T,\Pi_T)$, $(R,\Pi_R)$ and
$(\psi,\pi_{\psi})$ into the rescaled super-Hamilton (\ref{Htilde}) and the
supermomentum (\ref{H1tilde}), we obtain:
\begin{eqnarray}
  \label{resHtilde}
\tilde{\mathcal{H}}&=&R'\Pi_T+T'\Pi_R+\tfrac{1}{2}(R^{-1}\pi_{\psi}^2+R\psi'{^2}),\\  
  \label{resH1tilde}
\tilde{\mathcal{H}_1}&=&T'\Pi_T+R'\Pi_R+\psi'\pi_{\psi}.
\end{eqnarray}
Of course the constraints equations (\ref{tildecons}) and (\ref{tildecons1})
remain unaffected by the reparametrization, such that also for the reduced
action in the Einstein-Rosen coordinates the constraints equation read
\begin{eqnarray}
  \label{restildecons}
\tilde{\mathcal{H}}&=&0,\\
  \label{restildecons1}
\tilde{\mathcal{H}}_1&=&0.  
\end{eqnarray} 
The new canonical variables bring then the reduced ADM action in the
subsequent simplified form, whereby $S=S[T,\Pi_T,R,\Pi_R,\psi,\pi_{\psi}]$:
\begin{equation}
  \label{ERaction}
S=2\pi\int^{\infty}_{-\infty}dt\int^{\infty}_{0}dr\,(\Pi_{T}\dot{T}+\Pi_R\dot{R}+\pi_{\psi}\dot{\psi}-\tilde{\textrm{N}}\tilde{\mathcal{H}}-\tilde{\textrm{N}}^1\tilde{\mathcal{H}_1}).  
\end{equation}
By varying the action (\ref{ERaction}) with respect to $\tilde{\textrm{N}}$
and $\tilde{\textrm{N}}^1$ the constraint equations (\ref{restildecons}) and
(\ref{restildecons1}) are derived.

%%% Local Variables: 
%%% mode: latex
%%% TeX-master: "liz"
%%% End: 

\clearemptydoublepage
% Observables of the Einstein-Rosen Waves, bisobservables.tex
\chapter{Observables of the Einstein-Rosen Wave}
In this chapter we dedicate us to the main topic of this work. It is presented
how a set of phase space functions can be derived from the dynamical
variables, namely the scalar field $\psi$ and the canonically conjugate
momentum $\pi_{\psi}$. The phase space functions turn out to be observables of
the Einstein-Rosen waves. The condition the phase space functions have to
fulfill in order to be considered as observables will be discussed in
chapter six.\\
\indent Before we can look after the phase space functions, it is
necessary to express the canonically conjugate momentum $\pi_{\psi}$ as a
functional of the scalar field $\psi$ (\ref{gensolu}). 
\section{Canonical Momentum $\pi_{\psi}$}
The canonically conjugate momentum $\pi_{\psi}$ is derived following the
method presented in Torre's work \cite{torre91}. In the canonical formalism
the dynamic of a field is described by the Hamilton equation
\begin{equation}
  \label{hameq}
\dot{\psi}=\{\psi,\textrm{H}\},  
\end{equation}  
where the dot denotes the time derivative $\partial_t$. The scalar field
$\psi$ is well known, being the solution to the wave equation
(\ref{sfield}). The solution, we have found in chapter two, is 
\begin{equation}
  \label{scalarfiel}
  \psi(T,R)=\int^{\infty}_0 dk\,\textrm{J}_0 (kR)\left[\textrm{A}(k)\,\ex^{ikT}+\textrm{A}^{\ast}(k)\,\ex^{-ikT}\right].  
\end{equation}
The Hamilton function
\begin{equation}
  \label{linH}
\textrm{H}=\int^{\infty}_0 dr(\tilde{\textrm{N}}\tilde{\mathcal{H}}+\tilde{\textrm{N}}^1\tilde{\mathcal{H}}_1)  
\end{equation}
is given by the integral over the linear combination of the super-Hamiltonian $\tilde{\mathcal{H}}$
and the supermomentum $\tilde{\mathcal{H}}_1$, which were obtained for the reduced action of the
Einstein-Rosen spacetime.
\begin{eqnarray}
  \label{resHtilde_}
\tilde{\mathcal{H}}&=&R'\Pi_T+T'\Pi_R+\tfrac{1}{2}(R^{-1}\pi_{\psi}^2+R\psi'{^2}),\\  
  \label{resH1tilde_}
\tilde{\mathcal{H}_1}&=&T'\Pi_T+R'\Pi_R+\psi'\pi_{\psi}.
\end{eqnarray} 
The coordinates $T(r)$ and $R(r)$ are part of an embedding
$X^{\alpha}:\Sigma\rightarrow\mathbb{R}^4$, $\Sigma$ being the hypersurface at
$t=const.$, which maps the hypersurface $\Sigma$ into the flat spacetime. The
embeddings $X^{\alpha}(r)=(T(r),R(r),\varphi,z)$ are arbitrary cylindrically
symmetric slices on the interior of the spacetime and approach the
hypersurface $\textbf{T}=0$ for $r\rightarrow\infty$, where $\textbf{T}$ is
the Minkowskian time. The coordinates $T(r)$ and $R(r)$ do not depend on
$\varphi$ and $z$ due to the cylindrically symmetric model.\\
\indent If the Hamilton equation (\ref{hameq}) is solved then the canonically
conjugate momentum $\pi_{\psi}$ is obtained as a functional of the scalar
field $\psi$. In the following prearrangements are presented, which allow to
solve the Hamilton equation in a facilitated manner. It is then possible to
write the time derivative, denoted by a dot in (\ref{hameq}), by a functional
derivative of the embedding variables  $X^{\alpha}(r)=(T(r),R(r),\varphi,z)$.  
The definition of the unit vector on a given hypersurface is defined by
\begin{align}
  \label{c1}
X^{\alpha}_{\ ,a}\,n_{\alpha}&=0,\\  
  \label{c2}
g^{\alpha\beta}n_{\alpha}n_{\beta}&=-1,
\end{align}
where the index $a$ runs over $(1,2,3)$ labeling $(R,\varphi,z)$. The metric
$g_{\alpha\beta}$ is the Minkowskian metric written for the cylindrically
symmetric coordinates. The diagonal components are $(-1,1,R^2,1)$, while all
other components vanish. The matrix $X^{\alpha}_{\ ,a}$ reads
\begin{equation}
  \label{xmatrix}
X^{\alpha}_{\ ,a}=\left(\begin{array}{ccc}
T' & 0 & 0 \\
R' & 0 & 0 \\
0 & 1 & 0 \\
0 & 0 & 1
\end{array}\right). 
\end{equation}
From the condition (\ref{c1}) the two possible normal covectors on the
hypersurface $\Sigma$ are obtained:
\begin{align}
  \label{normalvec1}
n^{(1)}_{\alpha}&=(-R',T',0,0),\\
  \label{normalvec2}
n^{(2)}_{\alpha}&=(R',-T',0,0). 
\end{align}
For the further calculation we choose without restricting generality the first
one, which points out of the hypersurface in the negative time direction. The
normal covector is normalized on $-1$ by the second condition (\ref{c2}).
\begin{align}
\nn &n^{\alpha}=g^{\alpha\beta}n_{\beta}=(R',T',0,0),\\
\nn &A\,n^{\alpha}n_{\alpha}=A\,(-R'{^2}+T'{^2})=-1,\\
  \label{normfact}
    &\rightarrow\quad A=\frac{1}{R'{^2}-T'{^2}}.
\end{align}
For convenience we assign the normalizing factor $A$ to the covector, so to get in
the further calculation a more compact form for the momentum $\pi_{\psi}(r)$,
\begin{align}
  \label{unitcovec}
&n_{\alpha}=\frac{1}{R'{^2}-T'{^2}}\,(-R',T',0,0),\\
  \label{unitvec}
&n^{\alpha}=(R',T',0,0).
\end{align}
Hence, the induced metric on the hypersurface is given by the embedding
variables, 
\begin{equation}
  \label{indmet}
g_{ab}=X^{\alpha}_{\ ,a}\,g_{\alpha\beta}\,X^{\beta}_{\ ,b}.  
\end{equation}
The components in the diagonal of the matrix $g_{ab}$ read
\begin{equation}
  \label{diagind}
(R'{^2}-T'{^2},R^2,1),  
\end{equation}
all other components are zero. The projection operator $X^a_{\ \alpha}$, which takes a spacetime vector into
a vector in the hypersurface, is expressed by the induced metric,
\begin{equation}
  \label{projvec}
X^a_{\ \alpha}=g_{\alpha\beta}\,X^{\beta},_{\ b}\,g^{ba}= \left(\begin{array}{ccc}
\frac{-T'}{R'{^2}-T'{^2}} & 0 & 0 \\
\frac{R'}{R'{^2}-T'{^2}} & 0 & 0 \\
0 & 1 & 0 \\
0 & 0 & 1
\end{array}\right). 
\end{equation}
If the constraint functions (\ref{resHtilde_}) and (\ref{resH1tilde_}) are taken together to a
covector then it is possible to express the Hamilton functions as follows:
\begin{equation}
  \label{hamfct}
\textrm{H}=\int^{\infty}_0 dr\;\tilde{\textrm{N}}^{\alpha}\tilde{\mathcal{H}}_{\alpha},  
\end{equation}
where $\alpha=0,1,2,3$. Of course $\tilde{\mathcal{H}}_2=0=\tilde{\mathcal{H}}_3$ due to the
symmetries, as we have learned in section (\ref{secADMCSS}). In this notation
the constraint equations read: $\tilde{\mathcal{H}}_{\alpha}=0$. Then also the
Einstein-Rosen momenta $\Pi_T$ and $\Pi_R$ in (\ref{resHtilde_}) and
(\ref{resH1tilde_}) respectively canonically conjugate to the embeddings
$T(r)$ and $R(r)$ are taken together to a covector,
\begin{equation}
  \label{pialpha}
\Pi_{\alpha}=(\Pi_{T},\Pi_{R},0,0).
\end{equation}
Thus the Hamilton function in (\ref{hamfct}) is newly expressed by,
\begin{align}
  \label{newham}
&\tilde{\mathcal{H}}_{\alpha}=\Pi_{\alpha}-g^{-1/2}n_{\alpha}h+X^{a}_{\ \alpha}h_{a},\\  
\nn \\
  \label{h0}
&h=\tfrac{1}{2}\left(\pi_{\psi}^2+R^2\psi'{^2}\right),\\  
  \label{h1}
&h_a=\left(\pi_{\psi}\psi',0,0\right).
\end{align}
The constraint function in this form allows to solve the Hamilton
equation in a easier way.
\begin{equation}
  \label{easyham}
\dot{\psi}(t,r)=\left\{\psi(t,r),\int^{\infty}_0 dr'\;\tilde{\textrm{N}}^{\alpha}(r')\tilde{\mathcal{H}}_{\alpha}(r')\right\}.  
\end{equation}
In order to compute the Poisson bracket, it is necessary to choose a lapse and
a shift function. In our case the fixing of the $\tilde{\textrm{N}}^{\alpha}$ is not
needed, as the introduction of embeddings selects a precise folium (an instant
of time) i.e. $t=const$., what just corresponds to the choice of
$\tilde{\textrm{N}}^{\alpha}$. In fact, only the $r$-dependence of the quantities
will be noted, for example $\psi(r)$ instead of $\psi(T(t,r),R(t,r))$.
The choice of the embeddings permits to replace the time derivative by the functional derivative. 
\begin{equation}
  \label{funcderiv}
\frac{d}{dt}=\int^{\infty}_0
dr\left(\frac{d}{dt}X^{\alpha}(r)\right)\frac{\delta}{\delta
  X^{\alpha}(r)}=\int^{\infty}_0 dr\;\tilde{\textrm{N}}^{\alpha}(r)\,\frac{\delta}{\delta
  X^{\alpha}(r)},  
\end{equation}
where it has been utilized: $\frac{d}{dt}X^{\alpha}(t,r)= \tilde{\textrm{N}}^{\alpha}$. If
the operator (\ref{funcderiv}) is applied on the time derivative of the wave
function, one gets
\begin{equation}
  \label{timederiv}
\dot{\psi}(t,r)=\int^{\infty}_0 dr'\;\tilde{\textrm{N}}^{\alpha}(r')\,\frac{\delta\psi(r)}{\delta
  X^{\alpha}(r')}=\int^{\infty}_0 dr'\;\tilde{\textrm{N}}^{\alpha}(r')\left\{\psi(r),\tilde{\mathcal{H}}_{\alpha}(r')\right\}.  
\end{equation}
The right hand side is obtained by simplifying the same side of the
Hamilton equation (\ref{easyham}), considering that Poisson brackets are
linear in both arguments, so
\begin{align}
  \label{easyhambrack}
\{\psi,\textrm{H}\}=\left\{\psi(r),\int^{\infty}_0
  dr'\;\tilde{\textrm{N}}^{\alpha}(r')\tilde{\mathcal{H}}_{\alpha}(r')\right\}=\int^{\infty}_0 dr'\;\tilde{\textrm{N}}^{\alpha}(r')\{\psi(r),\tilde{\mathcal{H}}_{\alpha}(r')\}.
\end{align}
Comparing the integrands in (\ref{timederiv}), the following has to hold obviously.
\begin{equation}
  \label{defeq}
\frac{\delta\psi(r)}{\delta X^{\alpha}(r')}=\{\psi(r),\tilde{\mathcal{H}}_{\alpha}(r')\}.  
\end{equation}
If in the above equation the Poisson bracket is solved then the conditional
equation for the canonical momentum $\pi_{\psi}(r)$ comes out.
\begin{align}
  \label{condeq}
\nn \{\psi(r),\tilde{\mathcal{H}}_{\alpha}(r')\}&=\int^{\infty}_0 dr''\bigg(\frac{\delta\psi(r)}{\delta\psi(r'')}
\frac{\delta\tilde{\mathcal{H}}_{\alpha}(r')}{\delta\pi_{\psi}(r'')}-\frac{\delta\psi(r)}{\delta\pi_{\psi}(r'')}
\frac{\delta\tilde{\mathcal{H}}_{\alpha}(r')}{\delta\psi(r'')}\bigg)\\
\nn &=\int^{\infty}_0
dr''\;\delta(r-r'')\,\delta(r-r')\,[-g^{-1/2}n_{\alpha}\pi_{\psi}(r')+X^1_{\alpha}\psi'(r')]\\
&=-g^{-1/2}\;n_{\alpha}\;\pi_{\psi}(r')\;\delta(r-r')+X^1_{\alpha}\;\psi'(r')\;\delta(r-r').  
\end{align}
The canonical equation $\dot{\psi}=\{\psi,\textrm{H}\}$ has been reduced to 
\begin{equation}
  \label{canequ}
\frac{\delta\psi(r)}{\delta X^{\alpha}(r')}=\delta(r-r')\,[-g^{-1/2}n_{\alpha}\pi_{\psi}(r')+X^1_{\alpha}\psi'(r')]  
\end{equation}
To get the momentum, we set this form of the canonical equation in
(\ref{timederiv}), remembering $g^{-1/2}=R^{-1}$, $g$ is the determinant of the
Minkowskian metric in cylindrical coordinates.
\begin{align}
  \label{moment}
\nn \int^{\infty}_0 dr'\;\tilde{\textrm{N}}^{\alpha}(r')\,\frac{\delta\psi(r)}{\delta
     X^{\alpha}(r')}&=\int^{\infty}_0
     dr'\;\tilde{\textrm{N}}^{\alpha}(r')\;\delta(r-r')\,[X^1_{\alpha}\psi'(r')-R^{-1}n_{\alpha}\pi_{\psi}(r')]\\
\nn \int^{\infty}_0 dr'\;\tilde{\textrm{N}}^{\alpha}(r')\;\frac{\partial\psi(r)}{\partial
    X^{\alpha}(r')}\;\delta(r-r')&=\tilde{\textrm{N}}^{\alpha}(r)\,[X^1_{\alpha}\psi'(r)-R^{-1}n_{\alpha}\pi_{\psi}(r)]\\
    n^{\alpha}(r)\;\frac{\partial\psi(r)}{\partial
    X^{\alpha}(r)}&=n^{\alpha}(r)\;X^1_{\alpha}\psi'(r)-R^{-1}\;
    n^{\alpha}\,n_{\alpha}\;\pi_{\psi}(r).  
\end{align}
Since the smear function $\tilde{\textrm{N}}^{\alpha}(r)$ can be arbitrarily chosen,
it is allowed to substitute it with the unit vector $n^{\alpha}$. As $n^{\alpha}n_{\alpha}=-1$ and
$n^{\alpha}\,X^1_{\alpha}=0$, the formula for the momentum finally follows
from (\ref{moment}):
\begin{equation}
  \label{finalmoment}
\pi_{\psi}(r)=R(r)\,\big[R_{,r}\;\psi_{,T}\big(T(r),R(r)\big)+T_{,r}\;\psi_{,R}\big(T(r),R(r)\big)\big].  
\end{equation}
The canonical momentum $\pi_{\psi}$ is then given, as suited, by a functional
of the massless scalar field $\psi$. Inserting the derivatives with respect to
the embeddings the momentum is then written as a functional of the mode
expansion $\textrm{A}$ and $\textrm{A}^{\ast}$ contained in the field
$\psi$.
\section{Cauchy Data}\label{secCD}
The cylindrically symmetric wave equation (\ref{sfield}) as also equivalently
the Hamilton equation (\ref{hameq}) define the initial value problem for the
scalar field $\psi$. If initial data are given, also known as Cauchy data, the
dynamics of the field $\psi(T,R)$ is uniquely described by the differential
equation (\ref{sfield}). If the problem is solved in the canonical
formalism then an initial value of the field $\psi(T,R)$ and of the momentum
$\pi_{\psi}(T,R)$ must be given to uniquely fix the dynamics of the field by the
conditional equation of the field and momentum. The field $\psi(T,R)$ is known
as solution to the wave equation (\ref{sfield}), see for instance
(\ref{scalarfiel}). The momentum $\pi_{\psi}(T,R)$ is obtained by inserting (\ref{derivT})
and (\ref{derivR}) in equation (\ref{finalmoment}). Then the momentum
$\pi_{\psi}(T,R)$ reads:
\begin{align}
  \label{Pigen}
\nn \pi_{\psi}\left(T,R\right)&=iR\int^{\infty}_0
dk\,k\,\textrm{J}_0(kR)\left[\textrm{A}(k)\ex^{ikT}-\textrm{A}^{\ast}(k)\ex^{-ikT}\right]R,_r\\
&\ -R\int^{\infty}_0 dk\,k\,\textrm{J}_1(kR)\left[\textrm{A}(k)\ex^{ikT}+\textrm{A}^{\ast}(k)\ex^{-ikT}\right]T,_r.  
\end{align}
Now we affiliate the Cauchy data $\psi_0(r)$ and $\pi_{\psi_0}(r)$, which are
calculated for $T(r)=0$ and $R(r)=r$:
\begin{eqnarray}
  \label{inifield}
\psi_0(r)&\doteq&\psi(T,R)|_{T=0,R=r}=\int^{\infty}_0 dk\;\textrm{J}_0(kr)\left[\textrm{A}(k)+\textrm{A}^{\ast}(k)\right],\\  
     \label{iniPi}
\pi_{\psi_0}(r)&\doteq&\pi_{\psi}(T,R)|_{T=0,R=r}=ir\int^{\infty}_0 dk\;k\;\textrm{J}_0(kr)\left[\textrm{A}(k)-\textrm{A}^{\ast}(k)\right].
\end{eqnarray}
In order to enter the initial data $\psi_0(r)$ and $\pi_{\psi_0}(r)$ in the
general solution $\psi(T,R)$ and $\pi_{\psi}(T,R)$, $\psi_0(r)$ is adapted to
$\pi_{\psi_0}(r)$. Subsequently the sum of both is Hankel transformed.
\begin{eqnarray}
  \label{extraction}
\nn ikr\psi_0(r)&=&ir\int^{\infty}_0
dk\;k\;\textrm{J}_0(kr)\left[\textrm{A}(k)+\textrm{A}^{\ast}(k)\right]\\ 
\nn \pi_{\psi_0}(r)+ikr\psi_0(r)&=&2ir\int^{\infty}_0
dk\;k\;\textrm{J}_0(kr)\textrm{A}(k)\\
2i\textrm{A}(k)&=&\int^{\infty}_0 dr\;\textrm{J}_0(kr)\left[\pi_{\psi_0}(r)+ikr\psi_0(r)\right]. 
\end{eqnarray}
Hence, the expansion coefficients $\textrm{A}(k)$ and $\textrm{A}^{\ast}(k)$
are given as functionals of the Cauchy data $\psi_0(r)$ and $\pi_{\psi_0}(r)$
for $T(r)=0$ and $R(r)=r$.
\begin{eqnarray}
  \label{cid1}
\textrm{A}(k)&=&\tfrac{1}{2}\int^{\infty}_0 dr\;\textrm{J}_0(kr)\left[kr\,\psi_0(r)-i\,\pi_{\psi_0}(r)\right],\\  
  \label{cid2}
\textrm{A}^{\ast}(k)&=&\tfrac{1}{2}\int^{\infty}_0 dr\;\textrm{J}_0(kr)\left[kr\,\psi_0(r)+i\,\pi_{\psi_0}(r)\right].
\end{eqnarray}
By inserting the mode expansions (\ref{cid1}) and (\ref{cid2}) in the general
solution (\ref{scalarfiel}) and in the momentum (\ref{Pigen}), we managed to
express the canonical quantities $\psi(T,R)$ and $\pi_{\psi}(T,R)$ as
quantities of the Cauchy data $\psi_0(r)$ and $\pi_{\psi_0}(r)$.
\begin{align}
  \label{civppsi}
\nn  \psi\left[\textrm{A}(k),\textrm{A}^{\ast}(k);T(r),R(r)\right]
&=\int^{\infty}_0 dk\,\textrm{J}_0(kR)\bigg[\tfrac{1}{2}\int^{\infty}_0
  dr\;\textrm{J}_0(kr)\left(kr\,\psi_0-i\,\pi_{\psi_0}\right)\ex^{ikT}\\
&+\tfrac{1}{2}\int^{\infty}_0 dr\;\textrm{J}_0(kr)
  \left(kr\,\psi_0+i\,\pi_{\psi_0}\right)\bigg]\ex^{-ikT},
\end{align}
\begin{align}
  \label{civpPi}
\nn \pi_{\psi}\left[\textrm{A}(k),\textrm{A}^{\ast}(k);T(r),R(r)\right]
    &=iR\int^{\infty}_0 dk\;k\;\textrm{J}_0(kR)\bigg[\tfrac{1}{2}\int^{\infty}_0
      dr\;\textrm{J}_0(kr)\left(kr\,\psi_0-i\,\pi_{\psi_0}\right)\ex^{ikT}\\
\nn &\hspace{2.5cm}-\tfrac{1}{2}\int^{\infty}_0 dr\;\textrm{J}_0(kr)
      \left(kr\,\psi_0+i\,\pi_{\psi_0}\right)\ex^{-ikT}\bigg]R,_r\\
\nn &-R\int^{\infty}_0 dk\;k\;\textrm{J}_1(kR)\bigg[\tfrac{1}{2}\int^{\infty}_0
      dr\;\textrm{J}_0(kr)\left(kr\,\psi_0-i\,\pi_{\psi_0}\right)\ex^{ikT}\\
\nn &\hspace{2.5cm}+\tfrac{1}{2}\int^{\infty}_0 dr\;\textrm{J}_0(kr)\left(kr\,\psi_0+i\,\pi_{\psi_0}\right)\ex^{-ikT}\bigg]T,_r.\\
\end{align}
Hence, the dynamics is uniquely given by the initial data $\psi_0(r)$ and
$\pi_{\psi_0}(r)$ and so the initial value problem has been solved completely.
\section{Observables $\textrm{A}(k)$ and $\textrm{A}^{\ast}(k)$} 
In the final part of the present chapter we would like to deduce the phase
space functions $\textrm{A}(k)$ and $\textrm{A}^{\ast}(k)$, which turn
out to be observables of the cylindrically symmetric gravitational waves. By
deriving the field $\psi$ and its conjugate momentum $\pi_{\psi}$ as
functional of the Cauchy data (\ref{inifield}) and (\ref{iniPi}), we also
calculated the phase space functions $\textrm{A}(k)$ and
$\textrm{A}^{\ast}(k)$ for a specific hypersurface, namely for $T(r)=0$ and
$R(r)=r$. Of course, our goal is a generalization of these mode expansions, that is
to say for the phase space functions of an arbitrary hypersurface. To this end, we perform
some Hankel transformations of the canonical variables $\psi$ and $\pi_{\psi}$, whose sum finally will lead to the desired set of
observables.\\
\indent In the beginning we start to transform the momentum while multiplying
it with a phase,
\begin{align}
  \label{ht1}
\nn i\int^{\infty}_{0}dr\,\ex^{ikT}\textrm{J}_0(kR)\pi_{\psi}(r)
    =&-\ex^{ikT}\left[\textrm{A}(k)\ex^{ikT}-\textrm{A}^{\ast}(k)\ex^{-ikT}\right]\\
\nn  &-i\int^{\infty}_0 dr\,R\int^{\infty}_0 dk\,k\,\textrm{J}_0(k'R)\textrm{J}_1(kR)\\
     &\times\left[\textrm{A}(k)\ex^{ikT}+\textrm{A}^{\ast}(k)\ex^{-ikT}\right]T,_r\ex^{ikT}.
\end{align}
The field is treated in a similar way, in order to obtain the integral term in
the right hand side of the equation above.
\begin{align}
  \label{ht2}
\nn  i\int^{\infty}_{0}dr\,\ex^{ikT}\,k\,\textrm{J}_1(kR)T,_r\psi(r)&=i\int^{\infty}_{0}dr\,R\int^{\infty}_0 
     dk\,k\,\textrm{J}_0(k'R)\textrm{J}_1(kR)\\
     &\times\left[\textrm{A}(k)\ex^{ikT}+\textrm{A}^{\ast}(k)\ex^{-ikT}\right]T,_r\ex^{ikT}.
\end{align}
Further on $\psi(r)$ is transformed once more in order to get something
related to the first part of the transformation (\ref{ht1}).
\begin{equation}
  \label{ht3}
\int^{\infty}_{0}dr\,\ex^{ikT}\,k\,R\,R,r\,\textrm{J}_0(kR)\psi(r)=\ex^{ikT}\left[\textrm{A}(k)\ex^{ikT}+\textrm{A}^{\ast}(k)\ex^{-ikT}\right].  
\end{equation}
If the sum over the three transformations is taken, we easily obtain the
observables.
\begin{equation}
  \label{sumhts}
2\textrm{A}^{\ast}(k)=\int^{\infty}_{0}dr\,\ex^{ikT}\left\{\psi(r)\,k\,R\left[i\,T,_r\textrm{J}_1(kR)+R,_r\textrm{J}_0(kR)\right]+i\,\textrm{J}_0(kR)\pi_{\psi}\right\}.  
\end{equation}
Thus, the observables on an arbitrary hypersurface are:
\begin{eqnarray}
  \label{A}
\textrm{A}(k)&=&\tfrac{1}{2}\int^{\infty}_{0}dr\;\ex^{-ikT}\left\{\psi(r)\,k\,R\left[R,_r\textrm{J}_0(kR)-i\,T,_r\textrm{J}_1(kR)\right]-i\,\textrm{J}_0(kR)\pi_{\psi}\right\},\\  
  \label{Astar}
\textrm{A}^{\ast}(k)&=&\tfrac{1}{2}\int^{\infty}_{0}dr\;\ex^{ikT}\left\{\psi(r)\,k\,R\left[R,_r\textrm{J}_0(kR)+i\,T,_r\textrm{J}_1(kR)\right]+i\,\textrm{J}_0(kR)\pi_{\psi}\right\}.
\end{eqnarray}
These phase space functions yield just the mode expansions (\ref{cid1})
and (\ref{cid2}) if $T(r)=0$ and $R(r)=r$ are set in the observables above. So
(\ref{A}) and (\ref{Astar}) are really generalizations of the phase space functions (\ref{cid1})
and (\ref{cid2}).

%%% Local Variables: 
%%% mode: latex
%%% TeX-master: "liz"
%%% End: 

\clearemptydoublepage
%Gauge Invariance, bisgaugeinv.tex
\chapter{Algebra of the Observables}
For the developing of a quantum theory of the cylindrically symmetric
gravitational waves, observables of the system are needed, which enter the
theory as operators. So it is of crucial importance to investigate the gauge
invariant property of the phase space functions proposed to be observables of
the model. Then, if the derived phase space functions $\textrm{A}$ and
$\textrm{A}^{\ast}$ on an arbitrary hypersurface are suggested to be
observables of the treated model, gauge invariance of these functions has
to be assured. In the first part of this chapter we present the condition a phase space function has to
fulfill in order to be considered as an observable. Then we will show that the phase space
functions $\textrm{A}$ and $\textrm{A}^{\ast}$ are observables. In the
second part we will calculate the Poisson brackets of the observables in order
to derive their algebra.
\section{Gauge Invariant Phase Space Function}
A classical observable is a phase space function, which is gauge invariant on
the constraint surface \cite{henneaux94}. First class constraints $\gamma_a$
of a system are quantities, which are the generators
of infinitesimal gauge transformations. A phase space function, which remains
constant on the orbit in the constraint surface is gauge invariant. This is
equivalent to the vanishing Poisson brackets of the phase space function
$\textrm{F}$ and the first class constraints $\gamma_a$,
\begin{equation}
  \label{invar}
\{\textrm{F},\gamma_a\}\thickapprox 0.  
\end{equation}
The bracket is calculated on the whole phase space and afterwards its value is
taken on the constraint surface. The weak equality symbol '$\thickapprox$'
is to be understood in this way. In the following section an example for this
kind of calculation is given.
If (\ref{invar}) is fulfilled then we are allowed to consider the phase space function $\textrm{F}$
as an observable of the system.
\subsection{Gauge Invariance of $\textrm{A}(k)$ and $\textrm{A}^{\ast}(k)$}
In order to find out whether the mode coefficients $\textrm{A}(k)$ and
$\textrm{A}^{\ast}(k)$ of the cylindrically symmetric scalar wave are
gauge invariant, we have to assure that the Poisson bracket of these
functionals with the first class constraints vanishes on the constraint
surface, according to the formula (\ref{invar}). The first class constraints
are the super-Hamiltonian (\ref{resHtilde_}) and the supermomentum (\ref{resH1tilde_}).
Thus, $\textrm{A}(k)$, and therewith also $\textrm{A}^{\ast}(k)$, are observables if the following formula holds:
\begin{equation}
  \label{poissonAH}
\{\textrm{A}(k),\tilde{\mathcal{H}}_{\alpha}(r)\}\thickapprox 0.  
\end{equation}
The constraint submanifold $\Gamma$ is the surface where
$\tilde{\mathcal{H}}_{\alpha}=0$. Then, saying the Poisson bracket (\ref{poissonAH})
has to vanish on the constraint surface means that one evaluates firstly the
bracket and secondly, if still necessary, $\tilde{\mathcal{H}}_{\alpha}=0$ is set, so
to investigate whether the bracket really vanishes. The relation
(\ref{poissonAH}) can also be rewritten equivalently as,
\begin{equation}
  \label{eqpoissonAH}
\{\textrm{A}(k),\tilde{\mathcal{H}}_{\alpha}(r)\}|_{\Gamma:\tilde{\mathcal{H}}_{\alpha}=0}=0.  
\end{equation}
As shown in \cite{torre91} the Poisson bracket (\ref{poissonAH}) vanish, such
that the phase space function are observables. The calculation is performed for
a scalar field propagating in curved spacetime. The cylindrically symmetric
scalar field on a Minkowskian background is then a special case of the
presented computation. 
\section{Poisson Brackets}
In this section we are going to calculate the Poisson brackets of the phase
space functions $\textrm{A}(k)$ and $\textrm{A}^{\ast}(k)$. The motivation
therefore is to elaborate a Poisson algebra of these observables. To this
end we introduce the functions $\textrm{Q}(r)$ and
$\textrm{P}(r)$ written in \cite{torre91}, where just for consistency we retain our
normalization. 
\begin{eqnarray}
  \label{Q}
\textrm{Q}(r)&=&\int^{\infty}_0 dk\;\textrm{J}_0(kr)\left[\textrm{A}(k)+\textrm{A}^{\ast}(k)\right],\\  
  \label{P}
\textrm{P}(r)&=&i\,r\int^{\infty}_0 dk\,k\;\textrm{J}_0(kr)\left[\textrm{A}(k)-\textrm{A}^{\ast}(k)\right],
\end{eqnarray}
where $\textrm{Q}(r)=\psi_0(r)$ (\ref{inifield}) and
$\textrm{P}(r)=\pi_{\psi_0}(r)$ (\ref{iniPi}).
As we know from section \ref{secCD}, the functions $\textrm{Q}(r)$ and $\textrm{P}(r)$ are related to
$\textrm{A}(k)$ and $\textrm{A}^{\ast}(k)$ by a Hankel transformation, which
will be undone in the next step so to describe the phase space functions as
functionals of $\textrm{Q}(r)$ and $\textrm{P}(r)$.
\begin{eqnarray}
  \label{A-}
\textrm{A}(k)&=&\tfrac{1}{2}\int^{\infty}_0 dr\;\textrm{J}_0(kr)\left[k\,r\;\textrm{Q}(r)-i\;\textrm{P}(r)\right],\\   
  \label{A+} 
\textrm{A}^{\ast}(k)&=&\tfrac{1}{2}\int^{\infty}_0 dr\;\textrm{J}_0(kr)\left[k\,r\;\textrm{Q}(r)+i\;\textrm{P}(r)\right]. 
\end{eqnarray}
\subsection{Algebra of $\textrm{A}(k)$ and $\textrm{A}^{\ast}(k)$}
It is straight forward to compute the Poisson brackets for the observables
$\textrm{A}(k)$ and $\textrm{A}^{\ast}(k)$.
\begin{equation}
  \label{Abracket}
\left\{\textrm{A}(k),\textrm{A}^{\ast}(k')\right\}=\int^{\infty}_0
dr\left[\frac{\delta\textrm{A}(k)}{\delta\textrm{Q}(r)}
\frac{\delta\textrm{A}^{\ast}(k')}{\delta\textrm{P}(r)}-\frac{\delta\textrm{A}(k)}{\delta\textrm{P}(r)}
\frac{\delta\textrm{A}^{\ast}(k')}{\delta\textrm{Q}(r)}\right],  
\end{equation}
\begin{align}
  \label{deltaA}
\frac{\delta\textrm{A}(k)}{\delta\textrm{Q}(r)}&
=\tfrac{1}{2}kr\;\textrm{J}_0(kr),&
\frac{\delta\textrm{A}(k)}{\delta\textrm{P}(r)}&
=-\tfrac{i}{2}\;\textrm{J}_0(kr),\\
\nn \\  
  \label{deltaAstar}
\frac{\delta\textrm{A}^{\ast}(k')}{\delta\textrm{P}(r)}&
=\tfrac{i}{2}\;\textrm{J}_0(k'r),&
\frac{\delta\textrm{A}^{\ast}(k')}{\delta\textrm{Q}(r)}& =\tfrac{1}{2}k'r\;\textrm{J}_0(k'r).
\end{align}
Inserting the functional derivatives in (\ref{Abracket}), we obtain the
coefficients $\textrm{A}(k)$ and $\textrm{A}^{\ast}(k)$ to be observables,
which are canonical conjugate to each other.
\begin{align}
  \label{canconj1}
\nn \left\{\textrm{A}(k),\textrm{A}^{\ast}(k')\right\}&=\int^{\infty}_0
dr\left[\tfrac{i}{4}\,kr\;\textrm{J}_0(kr)\textrm{J}_0(k'r)+
\tfrac{i}{4}\,k'r\;\textrm{J}_0(k'r)\textrm{J}_0(kr)\right],\\
\nn &=\tfrac{i}{4}\,k\frac{1}{k}\delta(k-k')+\tfrac{i}{4}\,k'\frac{1}{k'}\delta(k-k'),\\
    &=\tfrac{i}{2}\,\delta(k-k').  
\end{align}
\begin{align}
  \label{canconj2}
\nn \left\{\textrm{A}(k),\textrm{A}(k')\right\}&=\int^{\infty}_0 dr\left[-\tfrac{i}{4}\,kr\;\textrm{J}_0(kr)\textrm{J}_0(k'r)+
\tfrac{i}{4}\,k'r\;\textrm{J}_0(k'r)\textrm{J}_0(kr)\right],\\
\nn &=-\tfrac{i}{4}\,k\frac{1}{k}\delta(k-k')+\tfrac{i}{4}\,k'\frac{1}{k'}\delta(k-k'),\\
&=0.  
\end{align}
\begin{align}
  \label{canconj3}
\nn \left\{\textrm{A}^{\ast}(k),\textrm{A}^{\ast}(k')\right\}&=\int^{\infty}_0 dr\left[\tfrac{i}{4}\,k'r\;\textrm{J}_0(kr)\textrm{J}_0(k'r)-\tfrac{i}{4}\,kr\;\textrm{J}_0(kr)\textrm{J}_0(k'r)\right],\\
\nn &=\tfrac{i}{4}\,k'\frac{1}{k'}\delta(k-k')-\tfrac{i}{4}\,k\frac{1}{k}\delta(k-k'),\\
&=0.  
\end{align}
Hence, the well-defined algebra for the phase space functions $\textrm{A}(k)$
and $\textrm{A}^{\ast}(k)$ of the cylindrical gravitational wave has been
elicited:
\begin{align}
  \label{algebra}
\left\{\textrm{A}(k),\textrm{A}^{\ast}(k')\right\}&=\tfrac{i}{2}\,\delta(k-k'),&
\left\{\textrm{A}(k),\textrm{A}(k')\right\}&=0,& \left\{\textrm{A}^{\ast}(k),\textrm{A}^{\ast}(k')\right\}&=0.
\end{align}
\subsection{Algebra of $\textrm{Q}(r)$ and $\textrm{P}(r)$}
Analogously, the Poisson brackets of $\textrm{Q}(r)$ and $\textrm{P}(r)$ can
be calculated, so to obtain the Poisson algebra of these observables. 
\begin{equation}
  \label{QP}
\left\{\textrm{Q}(r),\textrm{P}(r')\right\}=\int^{\infty}_0
dk\left[\frac{\delta\textrm{Q}(r)}{\delta\textrm{A}(k)}
\frac{\delta\textrm{P}(r')}{\delta\textrm{A}^{\ast}(k)}-\frac{\delta\textrm{Q}(r)}{\delta\textrm{A}^{\ast}(k)}
\frac{\delta\textrm{P}(r')}{\delta\textrm{A}(k)}\right],  
\end{equation}
\begin{align}
  \label{deltaQ}
\frac{\delta\textrm{Q}(r)}{\delta\textrm{A}(k)}&
=\textrm{J}_0(kr),&
\frac{\delta\textrm{Q}(r)}{\delta\textrm{A}^{\ast}(k)}&
=\textrm{J}_0(kr),\\
\nn \\  
  \label{deltaP}
\frac{\delta\textrm{P}(r')}{\delta\textrm{A}(k)}&
=i\,k\,r'\,\textrm{J}_0(kr'),&
\frac{\delta\textrm{P}(r')}{\delta\textrm{A}^{\ast}(r)}& =-i\,k\,r'\,\textrm{J}_0(kr').
\end{align}
From this it follows immediately the Poisson algebra for $\textrm{Q}(r)$ and $\textrm{P}(r)$:
\begin{align}
  \label{algebraQP}
\left\{\textrm{Q}(r),\textrm{P}(r')\right\}&=-2i\,\delta(r-r'),&
\left\{\textrm{Q}(r),\textrm{Q}(r')\right\}&=0,& \left\{\textrm{P}(r),\textrm{P}(r')\right\}&=0.  
\end{align}
It is not surprising to get the same Poisson algebra for $\textrm{A}(k)$ and
$\textrm{A}^{\ast}(k)$ as well as for $\textrm{Q}(r)$ and $\textrm{P}(r)$
until normalization and sign. The functions $\textrm{Q}(r)$ and
$\textrm{P}(r)$ are, from a mathematical point of view, just expressions for
the Hankel transformed real and imaginary parts of the
mode coefficients $\textrm{A}(k)$ and $\textrm{A}^{\ast}(k)$.
\section{Quantum Theoretical Property }
The Poisson algebra (\ref{algebra}) for the observables $\textrm{A}(k)$ and
$\textrm{A}^{\ast}(k)$ has an interesting feature if investigated from a
quantum theoretical point of view. As known the Poisson brackets multiplied by
the complex $i$ correspond to the commutators of quantum theory. Multiplying
the Poisson brackets in (\ref{algebra}) with the complex $i$, the commutation relations for the
creator and the annihilator operators of the scalar field - see for instant
\cite{sweinberg00} - are obtained up to a
normalizing factor $\sqrt{2}$. Hence, the
observables $\textrm{A}(k)$ and $\textrm{A}^{\ast}(k)$ can just be identified
with the creator and annihilator respectively.
\begin{eqnarray}
\nn  \label{cre}
\textrm{A}&\rightarrow&\textbf{A},\\
  \label{ann}
\textrm{A}^{\ast}&\rightarrow&\textbf{A}^{\dagger}.
\end{eqnarray}
Viewing now the cylindrically symmetric scalar field as a wave function
in quantum field theory and applying the operators (\ref{cre}) and (\ref{ann})
on the wave function,
the Fock space of the system can be constructed, which spans the Hilbert space
of the quantum states of the cylindrically symmetric scalar field.

%%% Local Variables: 
%%% mode: latex
%%% TeX-master: "liz"
%%% End: 

\clearemptydoublepage
% Conclusions, conclusion.tex

\chapter{Conclusion}
By a conformal deformation and a coordinate transformation into the inertial system the
general cylindrically symmetric metric is brought into diagonal form. The
introduction of the Einstein-Rosen coordinates allows a further simplification
of the metric. The symmetries and the special form of the metric reduce the
Einstein equations to a set of three differential equations. The master
equation is solved by the cylindrically symmetric massless scalar
field propagating on a Minkowskian background.\\ \newline
\indent The introduction of the reduced ADM action for the Einstein-Rosen wave, and therewith of the super-Hamilton and
supermomentum gives the opportunity to derive the
momentum, which is canonically conjugate to the scalar field, by solving the
Hamilton equation of it. Combining the scalar field with its canonically
conjugate momentum and performing several Hankel transformations it is
possible to isolate the mode expansions contained in the function of the
scalar field. Just these phase space functions turn out to be gauge
invariant and hence observables of the model. This is shown by investigating
the Poisson bracket of the phase space function and the first class
constraints. As it is required for observables the Poisson bracket
vanishes.\newline \\
\indent In order to guarantee the phase space functions to be canonical conjugate
to each other the Poisson brackets of them are computed. The result assure
the expected property and allow the correspondence of the observables
to the annihilator and creator operators in the formulation of the quantum
theory for the cylindrically symmetric scalar field on a Minkowskian
background. Acting on the scalar field, now considered as a wave
function, the Fock space can be developed, spanning therewith the Hilbert space
of the quantum states of the cylindrically symmetric scalar field.

%%% Local Variables: 
%%% mode: latex
%%% TeX-master: "liz"
%%% End: 

\clearemptydoublepage
\addcontentsline{toc}{chapter}{Acknowledgements}
\include{acknowledgements}
\clearemptydoublepage
\begin{appendix}
\clearemptydoublepage
% FORMELN UND GLEICHUNGEN im Appendix A, bisAappendix.tex .
% Appendix A, trisAppendix.tex
%
\chapter{Orthonormality Relation for the Bessel Function of the First Kind}
\section{One Dimensional Fourier Transformation\\in Cartesian Coordinates}
In the beginning we shall study the following integral:
\begin{equation}
\int^\infty_{-\infty}dk\;\ex^{-ikx}\;\ex^{ikx'}=\int^\infty_{-\infty}dk\;\ex^{i(x-x')k}.\label{intdiv}
\end{equation}
Unfortunately this integral is divergent! In order to obviate this trouble, a
suitable regularisation is carried out. To this end a new measure is defined,
which contains a proper damping, whereas $\epsilon>0$.
\begin{equation*}
[dk]\doteq(\theta(-k)\;\ex^{\epsilon k}+\theta(k)\;\ex^{-\epsilon k})\;dk.
\end{equation*}
Therewith the integral exists and can be calculated.
\begin{eqnarray}
\nn\int^\infty_{-\infty}[dk]\;\ex^{-i(x-x')k}&=&\int^0_{-\infty}dk\;\ex^{\epsilon
  k}\;\ex^{i(x-x')k}+\int^\infty_0dk\;\ex^{-\epsilon k}\;\ex^{i(x-x')k}\\
\nn &=&\frac{1}{\epsilon-i(x-x')}-\frac{1}{\epsilon+i(x-x')},\quad\mbox{with}\;\;u\doteq x-x'\\
\nn &=&\frac{2\epsilon}{\epsilon^2+u^2}.
\end{eqnarray}
The next step is to perform the limes $\epsilon\rightarrow 0$ to achieve  the
value of the initial integral (\ref{intdiv})\footnote{In \cite {gelfand60} we find:
 $\lim_{\epsilon\to 0}\;\frac{1}{\pi}\frac{\epsilon}{x^2+\epsilon
 ^2}=\delta(x)$, for $\epsilon>0$.}.
\beq
\int^\infty_{-\infty}dk\;\ex^{i(x-x')k}&=&
\nn\lim_{\epsilon\to 0}\;\int^\infty_{-\infty}[dk]\;\ex^{i(x-x')k}\\
&=&\lim_{\epsilon\to 0}\;\frac{2\epsilon}{\epsilon^2+u^2}= 2\pi\;\delta(x-x')\label{distribution} 
\eeq
If the distribution (\ref{distribution}) is applied on a test
function\footnote{$\mathcal{S}$ is the Schwartz space.} $\in\mathcal{S}$, so obviously the result is not
normalized. To adjust this non-conformance it is sufficient to assess the
integral (\ref{intdiv}) with $\frac{1}{2\pi}$.
Now we are allowed to undertake the identity map of the Fourier transformation
in one dimension.
Let's define the transformation and its inverse respectively as follows:
\beq
\nn \f (x)&=&\frac{1}{2\pi}\int^\infty_{-\infty}dk\;\ex^{ikx}\;\tilde{\f }(k)\\
\nn \tilde{\f }(k)&=&\int^\infty_{-\infty}dx'\;\ex^{-ikx'}\;\f (x')
\eeq           
The identity map reads
\begin{equation}
\label{idmap}
\f
(x)=\frac{1}{2\pi}\int^\infty_{-\infty}dx'\;\ex^{ikx}\int^\infty_{-\infty}dk\;\ex^{-ikx'}\f
(x'),
\end{equation}
whereby the integration order has been changed, since the integral over k is
regularizable, as shown above. So we obtain
\begin{eqnarray}
\label{calmap}
\nn \f
(x)&=&\frac{1}{2\pi}\int^\infty_{-\infty}dx'\int^\infty_{-\infty}dk\;\ex^{i(x-x')k}\;\f
(x')\\
&=&\frac{1}{2\pi}\int^\infty_{-\infty}dx'\;2\pi\;\delta(x-x')\;\f (x')=\f (x).
\end{eqnarray}
Finally, the identity map (\ref{idmap}) is normalized on one by the factor
$N_{d=1}=\frac{1}{2\pi}$.

\section{Two Dimensional Fourier Transformation\\in Cartesian Coordinates}
In order to understand better the switch from the transformation in Cartesian
coordinates to the one in cylindrical coordinates, first we shall
write down the relations in the first mentioned coordinates.\\ 
We take two vectors from the two dimensional vector space $\mathbb{R}^2$, namely
$\vec{x}\doteq(x_1,x_2)$ and $\vec{k}\doteq(k_1,k_2)$. The
Fourier transformation in $d=2$ and its inverse are defined respectively as follows:
\begin{eqnarray}
\label{ft2d1}
\f (\vec{x}) &\doteq& \frac{1}{(2\pi)^2}\int^\infty_{-\infty}
d^2k\;\ex^{i\vec{k}\vec{x}}\;\tilde{\f }(\vec{k})\\
\tilde{\f }(\vec{k}) &\doteq& \int^\infty_{-\infty}
d^2x'\;\ex^{-i\vec{k}\vec{x}'}\;\tilde{\f }(\vec{x}').\label{ft2d}
\end{eqnarray}
Applying the inverse on the transformation we get as is well known the
identity map\footnote{Belonging to (\ref{distribution}) the integral (\ref{intdiv}) is
regularizable. Thus, the integration order is arbitrary.}:
\begin{eqnarray}
\f (\vec{x})&=&\frac{1}{2\pi}\int^\infty_{-\infty}
d^2k\;\ex^{i\vec{k}\vec{x}}\frac{1}{2\pi}\int^\infty_{-\infty}
d^2x'\;\ex^{-i\vec{k}\vec{x}'}\;\tilde{\f }(\vec{x}')\label{idmap2d}\\
\nn &=&\frac{1}{(2\pi)^2}\int^\infty_{-\infty} d^2x'\;\int^\infty_{-\infty}
d^2k\;\ex^{i(\vec{x}-\vec{x}')\vec{k}}\;\tilde{\f }(\vec{x'})\\
\nn &=&\int^\infty_{-\infty}dx_1'\int^\infty_{-\infty}dx_2'\;\frac{1}{2\pi}\int^\infty_{-\infty}dk_1\;\ex^{i(x_1-x_1')k_1}\label{split}\\
& &\hspace{2.7cm}\times\;\frac{1}{2\pi}\int^\infty_{-\infty}dk_2\;\ex^{i(x_2-x_2')k_2}\;\tilde{\f
  }(x_1',x_2')\\
\nn &=&\frac{1}{(2\pi)^2}\int^\infty_{-\infty}d^2x'\;(2\pi)^2\;\delta^2(\vec{x}-\vec{x}')\tilde{\f
  }(\vec{x})\\
&=&\f (\vec{x}) \nonumber.
\end{eqnarray}
In the separated representation of (\ref{split}) we observe twice a one
dimensional integration over $k_1$ resp. $k_2$ of the form (\ref{intdiv}).
Belonging to (\ref{distribution}) each of them are equal to\linebreak
$2\pi\,\delta(x_i-x_i')$, $i=1,2$. Consequently, we understand the
reason why in (\ref{ft2d1}) the normalizing factor in two dimensions
$N_{d=2}=\frac{1}{(2\pi)^2}$ appears, analogously to the
Fourier-transformation in one dimension.  
\section{Hankel Transformation}
We start now with noting the two components of both vectors $\vec{x}\doteq(x_1,x_2)$ and $\vec{k}\doteq(k_1,k_2)$ in cylindrical
coordinates. That is 
\begin{equation*}
 x_1=x\cos\alpha, \quad x_2=x\sin\alpha \hspace{1.5cm} k_1=k\cos\beta,\quad k_2=k\sin\beta.
\end{equation*} 
For further need, we prepare a third vector from $\mathbb{R}^2$ in the same
coordinates: $\vec{x}'\doteq(x_1',x_2')$ with $ x_1'=x'\cos\alpha'$,
$x_2'=x'\sin\alpha'$. Thus, we get for the following products:
\begin{equation*}
\vec{k}\vec{x}=kx\cos(\alpha-\beta), \qquad \vec{k}\vec{x}'=kx'\cos(\alpha'-\beta). 
\end{equation*}
Please note that the scalars $k$, $x$ and $x'$ $\in \mathbb{R}$ are the absolute values of the
homonymic vectors. Furthermore the determinants of the Jacobi matrix for the integral over $x'$ and $k$
give the measures $dx'$ times $x'$ and $dk$ times $k$. A function
$\g (\vec{x})=\g (x_1,x_2)$ and its Fourier transformed function
$\tilde{\g}(\vec{x}')=\tilde{\g}(x_1',x_2')$ can be noted in polar coordinates as
follows:
\begin{align}
\label{polarfct}
&\g (\vec{x})=\h (x,\alpha)=\f (x)\ex^{in\alpha}\\  
&\tilde{\g}(\vec{x}')=\tilde{\h}(x',\alpha')=\tilde{\f}(x')\ex^{in\alpha'}.
\end{align}
If we write down the Fourier transformation identity map in cylindrical
coordinates, then it reads:
\begin{equation}
\label{idmapcyl}
\f (x)\ex^{in\alpha}=\frac{1}{(2\pi)^2}\int^\infty_0dk\,k\int^{\pi}_{-\pi}d\beta\,\ex^{ikx\cos(\beta-\alpha)}\int^\infty_0dx'\,x'\int^{\pi}_{-\pi}d\alpha'\,\ex^{-ikx'\cos(\alpha'-\beta)}\,\ex^{in\alpha'}\,\tilde{\f }(x') 
\end{equation}
We treat now the last equation in order to obtain finally the identity
(\ref{idmap2d}) containing the Bessel function $\textrm{J}_n(kx)$. For this purpose we prove that an integral of
the form 
\begin{displaymath}
\int^{2\pi}_0d\varphi\;\ex^{ikx\cos(\varphi-\varphi_0)}
\end{displaymath}
is equal to
\begin{displaymath}
\int^{2\pi}_0d\vartheta\;\ex^{ikx\cos\vartheta}.
\end{displaymath}
This is possible due to the symmetry property of the trigonometric functions if a whole period
is taken. That is gladly our case.
\begin{eqnarray}
\nn\int^{2\pi}_0d\varphi\;\ex^{ikx\cos(\varphi-\varphi_0)}&=&\int^{\varphi_0}_0d\varphi\;\ex^{ikx\cos(\varphi-\varphi_0)}+\int^{2\pi}_{\varphi_0}d\varphi\;\ex^{ikx\cos(\varphi-\varphi_0)}\\
\nn&=&\int^{2\pi}_{\varphi_0}d\varphi\;\ex^{ikx\cos(\varphi-\varphi_0)}+\int^{2\pi+\varphi_0}_{2\pi}d\varphi\;\ex^{ikx\cos(\varphi-\varphi_0)}\\
\nn&=&\int^{2\pi+\varphi_0}_{\varphi_0}d\varphi\;\ex^{ikx\cos(\varphi-\varphi_0)}\\
&=&\int^{2\pi}_0d\vartheta\;\ex^{ikx\cos\vartheta},\label{proofcos}
\end{eqnarray}
whereas $\vartheta\doteq\varphi-\varphi_0$.\\ 
\rightline{QED}
Thereafter we return to the crucial equation (\ref{idmapcyl}). We perform a
first substitution of variables, namely
\begin{equation}
\label{varcha1}
\beta-\alpha=\frac{\pi}{2}+\beta'.  
\end{equation}
The transformation map (\ref{idmapcyl}) changes to 
\begin{eqnarray}
\label{1chanmap}
\nn\f (x)\,\ex^{in\alpha}&=&\frac{1}{(2\pi)^2}\int^\infty_0dx'\,x'\int^\infty_0dk\,k\int^{\pi-\alpha}_{-\tfrac{3}{2}\pi-\alpha}d\beta'\,\ex^{-ikx\sin\beta'}\\
&\times&\int^{\pi}_{-\pi}d\alpha'\,\ex^{-ikx'\cos(\alpha'-\alpha-\beta-\frac{\pi}{2})}\,\ex^{in\alpha'}\,\tilde{\f }(x')  
\end{eqnarray}
In doing so we have employed the properties of the cosine function 
\begin{eqnarray}
\label{cosprop}
\cos(\varphi)&=&\cos(-\varphi)\\
\cos(\frac{\pi}{2}+\varphi)&=&-\sin(\varphi).  
\end{eqnarray}
In virtue of the periodicity of the trigonometric functions, the interval
$[-\tfrac{3}{2}\pi-\alpha,\frac{\pi}{2}-\alpha]$ can be traced back to the
original interval $[-\pi,\pi]$, cp. (\ref{proofcos}). Then a second
substitution is undertaken, namely
\begin{equation}
\label{varcha2}
\alpha'-\alpha-\beta'=\vartheta,  
\end{equation}
whereby one gets for the transformation newly\footnote{The integral succession can be switched
arbitrary, as the regularisation for the one dimensional case in section A.1 guarantees the existence of the integrals.}:
\begin{eqnarray}
\label{2chanmap}
\nn\f (x)\,\ex^{in\alpha}&=&\ex^{in\alpha}\int^\infty_0dx'\,x'\int^\infty_0dk\,k\,\tfrac{1}{(2\pi)}\int^{\pi}_{-\pi}d\beta'\,\ex^{-ikx\sin\beta'+in\beta'}\\
&\times&\tfrac{1}{(2\pi)}\int^{\pi}_{-\pi}d\vartheta\,\ex^{-ikx'\sin\vartheta+in\vartheta}\,\tilde{\f}(x').
\end{eqnarray}
Also for this step the periodicity of the trigonometric functions has been of
use, since one integrates at each case over a complete period. Further on the
identity
$\cos(\vartheta-\frac{\pi}{2})=\cos(\frac{\pi}{2}-\vartheta)=\sin\vartheta$ has
been utilized, too. The integral representation of the Bessel function $\textrm{J}_n(z)$ is given in \cite{smirnow55} by
\begin{equation}
\label{besseln} 
\textrm{J}_n(z)\doteq\frac{1}{2\pi}\int^\pi_{-\pi}d\varphi\;\ex^{i(n\varphi-z\sin\varphi)}.
\end{equation}
By setting $\vartheta=\varphi+\pi$ and making use of
$\cos(\varphi+\pi)=-\cos\varphi$ and $\sin(\varphi+\pi)=-\sin\varphi$, it can
be shown that the Bessel function $\textrm{J}_n(z)$ is a real function,
i.e. $\textrm{J}_n(z)=\textrm{J}^{\ast}_n(z)$. Comparing both integrals over
$\beta'$ and $\vartheta$ in (\ref{2chanmap}) with the integral representation
of the Bessel function $\textrm{J}_n(z)$, we remark that the integrals in
(\ref{2chanmap}) are just the Bessel functions. 
After latter cognitions the sums over $\beta'$ and $\vartheta$ can be
substituted with the Bessel function
$\textrm{J}_n(kx)$ and $\textrm{J}_n(kx')$,  replacing $z$ with $kx$
respectively with $kx'$. Thus, we get:
\begin{equation}
\label{idmapbessel}
\f (x)=\int^\infty_0dx'\;x'\int^\infty_0dk\;k\;\textrm{J}_n(kx)\textrm{J}_n(kx')\;\f (x')=\f (x)
\end{equation}                            
If we are interested in the Fourier transformation on the dependence of the
vector's $\vec{x}$ absolute value only, then the function f in (\ref{2chanmap}) is no more
subject to the angle $\alpha$. Recapitulating we put down:                                   
\begin{equation}
g(|\vec{x}|)=\textrm{f}(x)=\int^\infty_0dx'\;x'\int^\infty_0dk\;k\;\textrm{J}_n(kx)\textrm{J}_n(kx')\;\tilde{\f}(x')=\f(x).
\end{equation}
Analogously to the identity map in Cartesian coordinate (\ref{idmap2d}), the
Fourier transformation and its inverse, described in the new variables, can be obtained comparing just the form
in (\ref{idmap2d}) and the gained form in cylindrical coordinates in (\ref{idmapbessel}). The
so-called Hankel transformations read:
\beq
\nn \f (x)&=&\int^\infty_0dk\;k\;\textrm{J}_n(kx)\;\tilde{\f }(k)\\
\tilde{\f }(k)&=&\int^\infty_0dx'\;x'\;\textrm{J}_n(kx')\;\f (x').\label{hankel}
\eeq
These equations are called Hankel transformation. They are the analogous
expressions to the two dimensional Fourier transformation with cylindrical symmetry.
The normalization factor for this transformation is also given in the last two
equations, namely $N_{cyl. coord.}=1$. So we deduce that the Bessel function
defined in \cite{smirnow55} already are normalized. Observing properly (\ref{idmapcyl}),
we remark that the initial factor $\frac{1}{(2\pi)^2}$ has been absorbed in
both Bessel functions $\textrm{J}_n(kx)$, $\textrm{J}_n(kx')$ belonging to the definition
in \cite{smirnow55}.\\
\\
\indent As a main result of this appendix we want to write down explicitly the
normalization factor for the orthonormality relation of the Bessel function of
the first kind $\textrm{J}_n(kx)$. We can take the relation directly from the
expression (\ref{idmapbessel}): 
\begin{equation}
\label{idmapbesseltwo}
\f (x)=\int^\infty_0dx'\;x'\int^\infty_0dk\;k\;\textrm{J}_n(kx)\textrm{J}_n(kx')\;\f (x')=\f (x)
\end{equation}
The orthonormality relation is just the integral over $k$, which obviously has
to be equal to the Dirac $\delta$-function multiplied with the normalizing
factor.
\begin{equation}
 \label{ortho1}
\int^\infty_0dk\;k\;\textrm{J}_n(kx)\textrm{J}_n(kx')=\frac{1}{x'}\;\delta(x-x'), 
\end{equation}
with the normalization factor 
\begin{equation}
\label{norm2}
\mathcal{N}_{x'}\doteq\frac{1}{x'}.  
\end{equation}
For the Hankel transform $\tilde{f}(k)$ there is of course the analogous
orthonormality relation:
\begin{equation}
  \label{hankortho}
\int^\infty_0dx\;x\;\textrm{J}_n(kx)\textrm{J}_n(k'x)=\frac{1}{k'}\;\delta(k-k'),  
\end{equation}
with the corresponding normalization factor
\begin{equation}
  \label{hanknorm}
\mathcal{N}_{k'}\doteq\frac{1}{k'}.  
\end{equation}

%%% Local Variables: 
%%% mode: latex
%%% TeX-master: "liz"
%%% End: 

\clearemptydoublepage
% FORMELN UND GLEICHUNGEN im Appendix, bisappendix.tex .

\chapter{The Energy $\Gamma(T,R)$}
The metric (\ref{ermetric}) is determined among other things also by the function
$\Gamma(T,R)$. This function is given indirectly by the gradient formed by the
derivatives $\Gamma, _T(T,R)$ (\ref{currdens}) and $\Gamma, _R(T,R)$ (\ref{enerdens}), which describe
the energy density of the massless scalar field $\psi(T,R)$ on a Minkowsikian
spacetime background and respectively the radial energy current density. 
\begin{equation}
\label{gradient}
\vec{\nabla}\Gamma(T,R)=
\left(\begin{array}{c}
\Gamma,_T \\
\Gamma,_R 
\end{array} \right)
\end{equation}
In order to get the function $\Gamma(T,R)$ we have to integrate the gradient
field (\ref{gradient}). Starting from the point $\Gamma(0,0)$ we would like to
reach the arbitrary point $\Gamma(T,R)$ by choosing an arbitrary
path. However, at this place we make use of a fundamental property of gradient
fields. This sort of fields are conservative. This property implicates that
all path from the starting point $\Gamma(0,0)$ to the arbitrary point
$\Gamma(T,R)$ are equivalent. So we are free to choose a particular path
without restricting generality.
\begin{equation}
\label{path}
\Gamma(T,R)=\Gamma(0,0)+\int^T_0dT'\ \Gamma,_{T'}(T',0)+\int^R_0dR'\ \Gamma,_{R'}(T,R').
\end{equation}
We prove now that 
\begin{equation} 
\label{gammai}
\Gamma(T,R)=\int^R_0dR'\ \Gamma,_{R'}(T,R'),
\end{equation}
if we deal in the following with a pure radiative field without sources on the
axis of symmetry ($z$-axis).\\
\textit{Proof}: If we have no sources on the axis of symmetry, the spatial
geometry must be locally Euclidean on the mentioned axis. So the
circumference of a small circle $R$=const., $T$=const., $z$=const. is the
$2\pi$ multiple of its proper radius.
For $R$=const., $T$=const., $z$=const. the Einstein-Rosen line element 
\begin{equation}
\label{ERline}
g^{ER}_{\alpha\beta}(T,R)\,x^{\alpha}\,x^{\beta}=\ex^{\Gamma-\psi}\,\left(-dT^2+dR^2\right)+R^2\,\ex^{-\psi}\,d\varphi^2+\ex^{\psi}\,dz^2
\end{equation}
reduces to the formula
\begin{align}
\label{constline}
&ds^2=R^2\,\ex^{-\psi}\,d\varphi^2\\
\Rightarrow\ &ds^2=R\,\ex^{-\tfrac{1}{2}\psi}\,d\varphi.
\end{align}
As we have Euclidean space near the axis of symmetry, we can set the
radius $r=R\ \ex^{-\tfrac{1}{2}\psi}$, with $r$ the radius in Euclidean
geometry. The Euclidean circumference turns out to be 
\begin{equation}
\label{circum}
\textrm{C}=2\pi r=2\pi R\ \ex^{-\tfrac{1}{2}\psi}, 
\end{equation}
for a fixed value of $R$. In case that $R$ is variable, i.e. it is no more the
radius of the small circle around the $z$-axis, then the radius $r$ is given by
\begin{equation}
\label{radiuss}
r'=\int^R_0dR'\ \ex^{\tfrac{1}{2}(\Gamma -\psi)},
\end{equation}
for constant $\varphi$.
In this case the circumference C$'$ is different:
\begin{equation}
\label{radiusc}
\textrm{C}'=2\pi r'=2\pi\int^R_0dR'\ \ex^{\tfrac{1}{2}(\Gamma -\psi)}.
\end{equation}
Since the geometry near the $z$-axis ($R\to 0$) is Euclidean, because
there is no source, both circumferences has to be equal to each other.
\begin{equation}
\label{circumequal} 
\lim_{R\to 0}\ \frac{\textrm{C}}{\textrm{C}'}=1.
\end{equation}
And therefrom 
\begin{equation}
\label{radiusequal} 
\lim_{R\to 0}\ \frac{R\ \ex^{-\tfrac{1}{2}\psi}}{\int^R_0dR'\ \ex^{\tfrac{1}{2}\Gamma -\tfrac{1}{2}\psi}}=1.
\end{equation}
With the help of the method by ``de l'Hopital'', we get for the limes
\begin{equation} 
\label{limit}
\lim_{R\to 0}\ \frac{1+\tfrac{1}{2}R\frac{\partial}{\partial R}\psi
  (T,R)}{\ex^{\tfrac{1}{2}\Gamma (T,R)}}=1 \quad
\Rightarrow \frac{1}{\ex^{\tfrac{1}{2}\Gamma (T,0)}}=1.
\end{equation}
Obviously the last equation holds only if 
\begin{equation}
\label{gammazero}
\Gamma (T,0)=0, \quad\forall T\in\R.
\end{equation}
And especially it follows of course
\begin{equation}
\label{gammazerozero}
\Gamma (0,0)=0.
\end{equation}
The derivative of (\ref{gammazero}) with respect to the time $T$ is clearly
zero, too. Then the assertion has been proved:
\begin{equation}
\label{energy}
\Gamma (T,R)=\int^R_0dR'\ \Gamma,_{R'}(T,R').
\end{equation}
\rightline{QED}\\
Finally, we see that, since $\Gamma ,_R$ is the energy density, the function $\Gamma (T,R)$ describes the energy of the
massless scalar field $\psi$ (up to a factor $2\pi$) contained in a disk with radius $R$ and a
thickness of $\Delta z=1$ at
the time $T$.
\section{Computation of $\Gamma (T,R)$}
We evaluate now the energy $\Gamma (T,R)$ explicitly. Therefore we integrate
the left side of (\ref{energy}) with respect to the radius $R$. 
\begin{eqnarray}
\label{int}
\Gamma (T,R)=\int^R_0dR'\ \Gamma,_{R'}(T,R')=\tfrac{1}{2}\int^R_0dR'\ R'\
(\psi,_{\,T}^2+\psi,_{\,R'}^2),
\end{eqnarray}
with
\begin{equation}
\label{wave}
\psi(T,R)=\int^\infty_0dk\,k\;\mjo\left[a(k)\,\ex^{ikT}+a^{\ast}(k)\,\ex^{-ikT}\right]
\end{equation}
The dependence $\Gamma(T,R)$ on $R$ is due to the Bessel function in
$\psi(T,R)$. Indeed only the Bessel function is subject to the radius $R$. So the integration over $R$ concerns
this function only. In \cite{abramo72} there is the formula 11.3.29 for
integrating this kind of Bessel function product. The results are:
\begin{eqnarray}
\label{intbessel}
\nn \int^R_0dR'R'\;\mjos\;\mjois&=&\frac{R}{k^2-{k'}^2}\;\big[k\;\mjI\;\mjIi-k'\;\mjo\;\mjIi\big]\\
\nn \int^R_0dR'R'\;\mjIs\;\mjIis&=&\frac{R}{k^2-{k'}^2}\;\big[k\;\mjII\;\mjIi-k'\;\mjI\;\mjIIi\big]\\
\textsc{}
\end{eqnarray} 
Therewith the integration over $R$ has been performed, and the resulting large
expression can be cast in a compact form, if the following is set:
\begin{align}
\label{set}
\nn a_1(k) & \doteq a(k) & a_1(k') &
\doteq a(k')\\
a_2(k) & \doteq a^{\ast}(k) & a_2(k') & \doteq a^{\ast}(k').
\end{align}
The integration over $R$ produces eight addends. It is advantageous to notice that
four of them are nothing but the complex conjugate of the other four. Together
with the definitions in (\ref{set}), this property is helpful to cast the function
$\Gamma(T,R)$ in a very short form:
\begin{align}
\label{shortgamma}
\nn \Gamma(T,R)=\tfrac{1}{2}\int^\infty_0dk\int^\infty_0dk'\big[&\textrm{K}_{11}\ a_1(k)\ a_1(k')
+\textrm{K}^{\ast}_{11}\ a_2(k)\ a_2(k') \\ +&\textrm{K}_{12}\
a_1(k)\ a_2(k')+\textrm{K}^{\ast}_{12}\ a_2(k)\ a_1(k')\big],  
\end{align}
where
\begin{align}
\label{conju}
\nn \textrm{K}^{\ast}_{11}&=\textrm{K}_{22}\\
\textrm{K}^{\ast}_{12}&=\textrm{K}_{21},
\end{align}
with
\begin{align}
\label{shorts}
\nn \textrm{K}_{11}=\textrm{K}_{11}[k,k';T,R]=\ &k\ k'\ \frac{R}{k^2-{k'}^2}\ \ex^{-(k+k')T}\ \big[k\;\mjII\;\mjIi-k'\;\mjI\;\mjIIi \\
\nn -&k\;\mjI\;\mjoi+k'\;\mjo\;\mjIi\big],\\
\nn \\
\nn \textrm{K}_{12}=\textrm{K}_{12}[k,k';T,R]=\ &k\ k'\ \frac{R}{k^2-{k'}^2} \ \ex^{-(k-k')T}\ \big[k\;\mjI\;\mjoi-k'\;\mjo\;\mjIi \\
\nn +&k\;\mjII\;\mjIi-k'\;\mjI\;\mjIIi\big].
\end{align}

%%% Local Variables: 
%%% mode: latex
%%% TeX-master: "liz"
%%% End: 

\clearemptydoublepage
%Ricci tensor and Ricci scalar, Cappendix.tex.

\chapter{Ricci Tensor and Ricci Scalar of the Einstein-Rosen Metric}
In this appendix we present the calculation of the Ricci tensor and the Ricci
scalar of the Einstein-Rosen metric. The computation has been performed with
the help of the software MAPLE 6, in particular with a specific package,
``with(tensor)'', wherein the two tools ``RICCI'' (for Ricci tensor) and ``RS''
(for Ricci scalar) are contained. In the following two sections there are the results of both
computations. The motivation for this task are shown in the section 1.4.\\
The Einstein-Rosen metric (\ref{ermetric}), whereof the Ricci tensor and the Ricci scalar have
been elicited reads:
\begin{equation*}
g_{\alpha\beta}=
\left(\begin{array}{cccc}
-\ex^{\Gamma(T,R)-\psi(T,R)} & 0 & 0 & 0 \\
0 & \ex^{\Gamma(T,R)-\psi(T,R)} & 0 & 0 \\
0 & 0 & R^2\ex^{-\psi} & 0 \\
0 & 0 & 0 & \ex^{\psi}
\end{array}\right)
\end{equation*}
\section{Ricci Tensor}
By virtue of the symmetry of the Einstein-Rosen metric the Ricci tensor has
the following five non-vanishing components:
\begin{align}
  \label{ricci00}
  &\textrm{R}_{00}=-\frac{1}{2R}\left[R\;\Gamma,_{RR}-R\;\psi,_{RR}-R\;\Gamma,_{TT}+R\;\psi,_{TT}+\Gamma,_{R}-\psi,_{R}-R\;\psi,_{T}^2\right]\\
  \label{ricci01}
  &\textrm{R}_{01}=-\tfrac{1}{2}\left[\frac{\Gamma,_{T}}{R}-\;\psi,_{T}\;\psi,_{R}\right]\\
  \label{ricci10}
  &\textrm{R}_{10}=-\tfrac{1}{2}\left[\frac{\Gamma,_{T}}{R}-\;\psi,_{T}\;\psi,_{R}\right]\\
  \label{ricci11}
  &\textrm{R}_{11}=-\frac{1}{2R}\left[R\;\psi,_{RR}-R\;\Gamma,_{RR}+R\;\Gamma,_{TT}-R\;\psi,_{TT}+\psi,_{R}+\Gamma,_{R}-R\;\psi,_{R}^2\right]
\end{align}
\begin{align}  
  \label{ricci22}
  &\textrm{R}_{22}=-\tfrac{1}{2}R\;\ex^{-\Gamma}\left[R\;\psi,_{RR}+\psi,_{R}-R\;\psi,_{TT}\right]\\
  \label{ricci33}
  &\textrm{R}_{33}=\frac{1}{2R}\;\ex^{2\psi-\Gamma}\left[R\;\psi,_{RR}-R\;\psi,_{TT}+\psi,_{R}\right].
\end{align}
We remark that the two components $\textrm{R}_{01}$ and $\textrm{R}_{10}$ are equal to each
other. This is rather what we expected, since the Ricci tensor is
symmetric in the two indices.

\section{Ricci Scalar}
For the Ricci scalar we obtained the following expression:
\begin{equation}
\label{ricciscalar}
\textrm{R}=-\frac{1}{2R}\;\ex^{-\Gamma+\psi}\,\left[2R\;\Gamma,_{TT}-2R\;\Gamma,_{RR}+2R\;\psi,_{RR}-2R\;\psi,_{TT}+2\;\psi,_{R}+R\;\psi,_{T}^2-R\;\psi,_{R}^2\right].  
\end{equation}

%%% Local Variables: 
%%% mode: latex
%%% TeX-master: "liz"
%%% End: 

\end{appendix}
\clearemptydoublepage
\addcontentsline{toc}{chapter}{Bibliography}
%Anfang Bibliographie
\bibliographystyle{phaip}
\bibliography{biblio}
%Ende Bibliographie

%Ende Dokument
\end{document}